\documentclass[apj]{emulateapj}
\usepackage{amsmath}
\usepackage{natbib}

\begin{document}

\newcommand{\kms}{km s$^{-1}$}
\newcommand{\comdlimit}{$30$}
\newcommand{\cmhigh}{$100$}

\title{Line-of-Sight Structure Toward Strong Lensing Galaxy Clusters$^{*}$}

\author{Matthew B. Bayliss\altaffilmark{1,2},
Traci Johnson\altaffilmark{3},
Michael D. Gladders\altaffilmark{4,5},
Keren Sharon\altaffilmark{3} and 
Masamune Oguri\altaffilmark{6,7}
}

\email{mbayliss@cfa.harvard.edu}

\altaffiltext{*}{From observations taken with the MMT Observatory, a joint 
facility of the Smithsonian Institution and the University of Arizona; the 6.5 meter 
Magellan Telescopes located at Las Campanas Observatory, Chile; with 
the Gemini Observatory, which is operated by the Association of Universities 
for Research in Astronomy, Inc., under a cooperative agreement with the NSF 
on behalf of the Gemini partnership: The United States, Canada, Chile, Australia, 
Brazil and Argentina; and with the Subaru Telescope, which is operated by the 
National Astronomical Observatory of Japan}

\altaffiltext{1}{Department of Physics, Harvard University, 17 Oxford St., 
Cambridge, MA 02138}
\altaffiltext{2}{Harvard-Smithsonian Center for Astrophysics, 60 Garden St., 
Cambridge, MA 02138}
\altaffiltext{3}{Department of Astronomy, the University of Michigan, 500 Church St. 
Ann Arbor, MI 48109}
\altaffiltext{4}{Department of Astronomy \& Astrophysics, 
University of Chicago, 5640 South Ellis Avenue, Chicago, IL 60637}
\altaffiltext{5}{Kavli Institute for Cosmological Physics, 
University of Chicago, 5640 South Ellis Avenue, Chicago, IL 60637}
\altaffiltext{6}{Department of Physics, University of Tokyo, Tokyo 113-0033, Japan}
\altaffiltext{7}{Kavli Institute for the Physics and Mathematics of the Universe 
(Kavli IPMU, WPI), University of Tokyo, Chiba 277-8583, Japan}

\begin{abstract}

We present an analysis of the line-of-sight structure toward a sample of ten 
strong lensing cluster cores. Structure is traced by groups that are identified 
spectroscopically in the redshift range, 0.1 $\leq$ z $\leq$ 0.9, and we 
measure the projected angular and comoving separations between each 
group and the primary strong lensing clusters in each corresponding line of 
sight. From these data we measure the distribution of projected angular 
separations between the primary strong lensing clusters and uncorrelated 
large scale structure as traced by groups. We then compare the observed 
distribution of angular separations for our strong lensing selected lines of 
sight against the distribution of groups that is predicted for clusters lying 
along random lines of sight. There is clear evidence for an excess of structure 
along the line of sight at small angular separations ($\theta \leq 6\arcmin$) 
along the strong lensing selected lines of sight, indicating that uncorrelated 
structure is a significant systematic that contributes to producing galaxy clusters 
with large cross sections for strong lensing. The prevalence of line-of-sight 
structure is one of several biases in strong lensing clusters that can potentially 
be folded into cosmological measurements using galaxy cluster samples. 
These results also have implications for current and future studies -- such 
as the Hubble Space Telescope Frontier Fields -- that make use of massive 
galaxy cluster lenses as precision cosmological telescopes; it is essential 
that the contribution of line-of-sight structure be carefully accounted for in 
the strong lens modeling of the cluster lenses.
\end{abstract}

\keywords{galaxies: clusters: strong lensing --- galaxies: distances and redshifts 
--- techniques: spectroscopic --- large-scale structure of universe}

\section{Introduction}
\label{sec:intro}

Strong lensing galaxy clusters are a small and extreme subset of the general 
cluster population, and as such, cluster lenses provide a valuable tracer of the 
rarest and most over-dense regions in the cosmic web. The abundance of galaxy 
cluster-scale strong lenses can be compared against predictions of the global 
strong lensing efficiency of galaxy clusters in simulations to test the concordance 
cosmological paradigm. The typical measurement that is made is the number 
density of giant arcs that are observed around samples of massive galaxy clusters 
\citep[``giant arc statistics'', see the review by][]{meneghetti2013}. Because 
cluster-scale lenses are rare these comparisons have historically been limited 
to using very small observational samples of arcs, but the results consistently 
find a significant excess of arcs observed compared to predictions from simulations \citep{grossman1988,bartelmann1998,cooray1999,luppino1999,zaritsky2003,
gladders2003,li2006,hennawi2007,horesh2011,meneghetti2011}. The notable 
difference between the frequency of giant arcs forming around massive clusters 
has been labeled the ``giant arc statistics problem'' for more than a decade 
\citep{bartelmann1998}. 

Reconciling the difference between the abundance of giant arcs predicted and 
observed is an outstanding problem in observational cosmology, and 
solving this problem requires an improved understanding of the detailed astrophysics 
that contribute to the global strong lensing properties of massive clusters. The physical 
processes that are responsible for making a small fraction of massive clusters into 
efficient gravitational lenses are not fully understood. Understanding these processes 
is also important because galaxy cluster strong lenses provides us with the best 
``natural telescopes'' for studying the distant universe, and several large observational 
initiatives are underway \citep[CLASH;][]{postman2012} or planned (e.g., the Hubble Space 
Telescope Frontier Fields\footnote{http://www.stsci.edu/hst/campaigns/frontier-fields/}) 
that will rely on using precision strong lens modeling of the magnification by strong lensing 
galaxy clusters to study the background universe.

\begin{deluxetable*}{lccrccc}[h]
\tablecaption{Spectroscopic Observations\label{targettable}}
\tablewidth{0pt}
\tabletypesize{\tiny}
\tablehead{
\colhead{Cluster Name} &
\colhead{ $\alpha$ (J2000.0) } &
\colhead{ $\delta$ (J2000.0) } &
\colhead{UT Date}  &
\colhead{Instrument} &
\colhead{Disperser} &
\colhead{Filter} 
}
\startdata
%
Abell 1703 &  13 15 05.24  & $+$51 49 02.6  & Mar 17 2012 & 
MMT/Hectospec  &  270 l/mm Grating  & -- \\
SDSS J0851+3331 &  08 51 38.86  &  $+$33 31 06.1  & Feb 19 2012  & 
MMT/Hectospec  &  270 l/mm Grating  & --  \\
SDSS J0915+3826 & 09 15 39.00 & $+$38 26 58.5 & Feb 19 2012  & 
MMT/Hectospec  &  270 l/mm Grating  & --  \\
SDSS~J0957+0509  &  09 57 38.50 & $+$05 09 21.6   & Jan 27 2009 & 
Magellan/IMACS+GISMO  & 150 l/mm Grating  & WB4800-7800  \\   
SDSS~J0957+0509  &  09 57 37.34 & $+$05 09 49.9   & Apr 09 2013 &  
Magellan/IMACS f/2 &  200 l/mm Grism  & --    \\  
SDSS J1038+4849 &  10 38 42.90 & $+$48 49 18.7  & Mar 16 2012  & 
MMT/Hectospec  &  270 l/mm Grating  & --  \\
SDSS J1050+0017\tablenotemark{a} &  10 50 40.20 & $+$00 16 17.6  & Mar 17 2013  & 
Magellan/IMACS f/2  &  200 l/mm Grism  & --  \\
SDSS J1050+0017\tablenotemark{a} &  10 50 40.20 & $+$00 16 17.6  & Mar 17 2013  & 
Magellan/IMACS f/2  &  200 l/mm Grism  & --  \\
SDSS J1050+0017 &  10 50 41.83    &  $+$00 17 18.1    & Mar 29 2012  & 
Gemini/GMOS North  &  R400 Grating  & OG515  \\
SDSS J1152+3313 &  11 52 00.15 &  $+$33 13 42.1  & Mar 16 2012  & 
MMT/Hectospec  &  270 l/mm Grating  & --  \\
SDSS J1226+2149 &  12 26 51.11 & $+$21 49 52.3  & Feb 19 2012  & 
MMT/Hectospec  &  270 l/mm Grating  & --  \\
SDSS J1226+2149\tablenotemark{b} &  12 26 50.70 &  $+$21 52 38.4   & Apr 20 2009 & 
Magellan/IMACS+GISMO  &  150 l/mm Grating  & WB4800-7800  \\
SDSS J1226+2149 &  12 26 50.42 & $+$21 49 53.0   & Apr 21 2009 & 
Magellan/IMACS+GISMO  &  150 l/mm Grating  & WB4800-7800  \\
SDSS J1226+2149\tablenotemark{b} & 12 26 50.70 &  $+$21 52 38.4 & May 27 2009 & 
Magellan/IMACS+GISMO  &  300 l/mm Grating  & WB4300-6750   \\
SDSS~J1329+2243  &  13 29 36.54  &  $+$22 43 16.7 & Jun 02 2011 & 
Gemini/GMOS North  &  R400 Grating  & OG515  \\
SDSS~J1329+2243  & 13 29 34.50  & $+$22 43 16.2  & Apr 09 2013 &  
Magellan/IMACS f/2 &  200 l/mm Grism  & WB3800-7000  
\enddata
\tablenotetext{a}{~SDSS J1050+0017 was observed with two different mutli-slit masks on March 17, 2012.}
\tablenotetext{b}{~One GISMO mask for SDSS J1226+2149 was re-observed in May 2009 due to poor weather during the exposures in Apr 2009.}
\end{deluxetable*}

A variety of different astrophysical effects can be invoked to explain how  
cluster lenses come to have large strong lensing cross sections, and to alleviate 
the tension between measurements of giant arc statistics and predictions from 
simulations. In additional to understanding line-of-sight structure, potentially important 
factors include accounting for baryons in the form of central massive galaxies and 
substructure \citep{flores2000,meneghetti2003,hennawi2007,
meneghetti2010}, cooling baryons dragging dark matter into cluster cores 
\citep{puchwein2005,rozo2008,wambsganss2008,blanchard2013}, major mergers 
\citep{torri2004,fedeli2006,redlich2012}, and the redshift distribution of the 
background galaxy source population \citep{hamana1997,oguri2003,
wambsganss2004,bayliss2011a,bayliss2012}, and also more exotic 
cosmological explanations such as primordial non-gaussianity 
\citep{daloisio2011a}.

Each of the factors described above focuses on the properties in the lens or source 
planes, where the typical implicit assumption in the description of a strong lensing 
system is that the lensing potential of a strong lens is concentrated in a single region 
with a size that is much smaller than the distances separating the observer/lens/source 
-- i.e., a single virialized structure such as a galaxy group or cluster. This simplifying 
assumption is convenient in that it confines the deflections due to gravitational lensing 
to a single plane, but neglects deflections due to other intervening mass distributions 
along the line of sight between the observer and the source. Furthermore, studies of 
simulated halos indicate that line-of-sight structure can introduce non-negligible 
systematic uncertainties in lensing-based measurements \citep{dalal2005,king2007}.

Ray tracing in simulations provides a range of results regarding the contribution of 
line-of-sight structure to galaxy cluster scale strong lenses \citep{wambsganss2005,
hilbert2007,puchwein2009,daloisio2011b,daloisio2013}, but observational constraints 
are so far nonexistent. There are individual examples in the literature of galaxy-scale 
lenses that receive boosts to their strong lensing cross sections due to intervening 
structure -- typically galaxy groups and clusters alone the line of sight 
\citep{fassnacht2006,wong2011}, as well as studies of the statistical relationship 
between galaxy-scale lenses and large scale structure tracers such as galaxy 
environmental density \citep{faure2009,fassnacht2011}. These studies focus on 
smaller gravitational lenses (e.g., small Einstein radius, r$_{E} \lesssim$ 3.5\arcsec); 
there is a notable lack of work exploring the importance of line-of-sight structure in 
real sample of galaxy cluster scale strong lenses, i.e., the most powerful gravitational 
lenses in the universe. In this paper we examine the line-of-sight structure toward a 
sample of galaxy clusters selected specifically for their strong lensing properties and 
high magnifications.

This paper is organized as follows: in $\S$~\ref{sec:obs} we describe the origin and 
reduction of the spectroscopic data sets that inform our analyses. In 
$\S$~\ref{sec:redshift_analysis} we identify spectroscopic members of the primary 
lensing clusters, as well as over-densities in redshift space that indicate the presence 
of likely projected structures along the line of sight. In $\S$~\ref{sec:discussion} we 
discuss the constraints that our observations provide on the frequency with which 
uncorrelated structures contribute toward the strong lensing cross section of massive 
galaxy clusters, and in $\S$~\ref{sec:conc} we summarize our results and their 
implications. All cosmological calculations performed for this paper assume a 
standard flat $\Lambda$ cold dark matter ($\Lambda$CDM) cosmology with 
$H_{0} = 73$ km s$^{-1}$ Mpc$^{-1}$, and matter density $\Omega_{M} = 0.25$. 
Magnitudes presented in this paper are in the AB system, calibrated against the 
SDSS. 


\section{Cluster Sample and Observations}
\label{sec:obs}

\begin{deluxetable*}{lcccccccccc}[t]
\tablecaption{Example Galaxy Redshift Sample\label{redshifttable}}
\tablewidth{0pt}
\tabletypesize{\tiny}
\tablehead{
\colhead{Galaxy ID} &
\colhead{ $\alpha$ {\tiny (J2000.0)} } &
\colhead{ $\delta$ {\tiny (J2000.0)} } &
\colhead{$g$ mag} &
\colhead{$\sigma_{g}$} &
\colhead{$r$ mag} &
\colhead{$\sigma_{r}$} &
\colhead{$i$ mag} &
\colhead{$\sigma_{i}$} &
\colhead{redshift} &
 \colhead{$\sigma_{redshift}$}
 }
\startdata
J131511.0+514653 & 13 15 11.07 & +51 46 53.7 &  23.17 &   0.03 &  21.80 &   0.03 &  21.32 &   0.03 &  0.26900 &  0.00060  \\
J131502.2+514951 & 13 15 02.26 & +51 49 51.2 &  22.99 &   0.03 &  21.23 &   0.03 &  20.82 &   0.03 &  0.27070 &  0.00060  \\
J131458.0+514916 & 13 14 58.10 & +51 49 16.3 &  22.38 &   0.03 &  20.93 &   0.03 &  20.42 &   0.03 &  0.28860 &  0.00060  \\
J131505.0+514606 & 13 15 05.04 & +51 46 06.3 &  23.39 &   0.03 &  22.58 &   0.03 &  22.24 &   0.03 &  0.29090 &  0.00050  \\
J131504.2+514750 & 13 15 04.26 & +51 47 50.8 &  24.66 &   0.04 &  23.42 &   0.03 &  23.03 &   0.03 &  0.28000 &  0.00100  \\
J131509.0+514622 & 13 15 09.01 & +51 46 22.8 &  23.08 &   0.03 &  21.79 &   0.03 &  21.35 &   0.03 &  0.27050 &  0.00050  \\
J131511.1+514557 & 13 15 11.12 & +51 45 57.7 &  22.78 &   0.03 &  21.49 &   0.03 &  21.03 &   0.03 &  0.27660 &  0.00050  \\
J131508.8+514545 & 13 15 08.89 & +51 45 45.5 &  23.73 &   0.03 &  22.01 &   0.03 &  21.85 &   0.03 &  0.27810 &  0.00050  \\
J131505.3+514536 & 13 15 05.37 & +51 45 36.8 &  22.80 &   0.03 &  22.38 &   0.03 &  22.09 &   0.03 &  0.27660 &  0.00050  \\
J131506.3+515428 & 13 15 06.39 & +51 54 28.1 &  19.62 &   0.03 &  18.07 &   0.03 &  17.54 &   0.03 &  0.27400 &  0.00100 
\enddata
\tablenotetext{}{~All magnitudes are AB.}
\end{deluxetable*}

\subsection{The Sloan Giant Arcs Cluster Lens Sample}
\label{sec:sample}

Incidents of strong lensing are identifiable by the formation of giant arcs, 
which are multiply imaged background sources that have been highly 
magnified by a foreground gravitational potential. The galaxy clusters studied 
in this work are drawn from the Sloan Giant Arcs Survey (SGAS; Gladders et al. 
in prep); numerous SGAS lenses, including the strong lensing systems studied here, 
have been previously published as used for various astrophysical and cosmological 
analyses \citep{oguri2009a,koester2010,bayliss2010,bayliss2011a,bayliss2011b,
bayliss2012,oguri2012,dahle2013,gladders2013,blanchard2013,bayliss2013b}. 
We refer the reader to those papers, especially \citet{bayliss2011b}, for further 
information on the SGAS sample and how it was defined. The strong and weak 
lensing properties of the clusters analyzed here are all presented in detail in 
\citet{oguri2012}; the entire sample has Einstein radii $\theta_{E} >$ 
5\arcsec, and eight of the nine have $\theta_{E} \geq$ 9\arcsec.

The subset of SGAS cluster lenses we focus on in this paper are those systems 
for which we have extensive spectroscopic followup (i.e., $\gtrsim$170 redshifts) in 
the field centered on the primary galaxy cluster lens. In practice this selection stems 
from observations conducted in a pseudo-queue mode at the MMT observatory; 
the clusters discussed here were all part of a list of 25 SGAS clusters that were 
proposed for MMT/Hectospec observations. Unfortunately 
we were only able to observe six clusters with the MMT. The Hectospec data 
are supplemented by three clusters which where then observed from Magellan with 
IMACS (essentially those clusters far enough south to be observable from Las 
Campanas Observatory). The selection of the sample analyzed in this paper is 
independent of the known physical properties of the lenses (e.g., redshift, mass). 
All new spectroscopic observations presented in this paper are summarized in 
Table~\ref{targettable}, and are supplemented by the available literature 
redshifts described in $\S$~\ref{sec:literature}.

\begin{figure*}[t]
\centering
\includegraphics[scale=0.35]{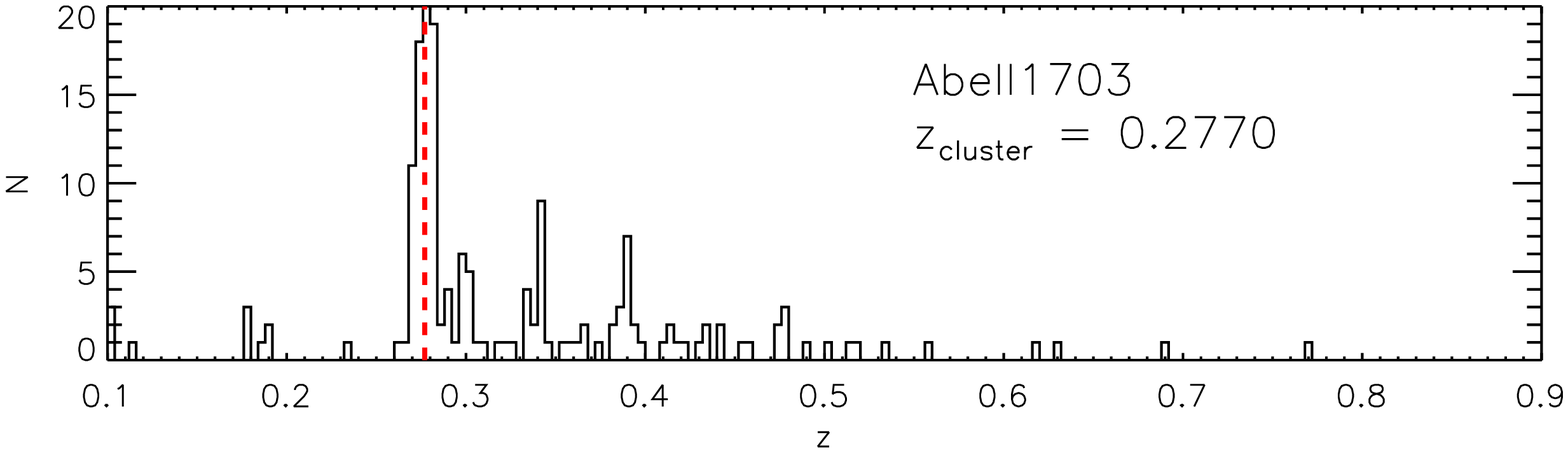}
\includegraphics[scale=0.35]{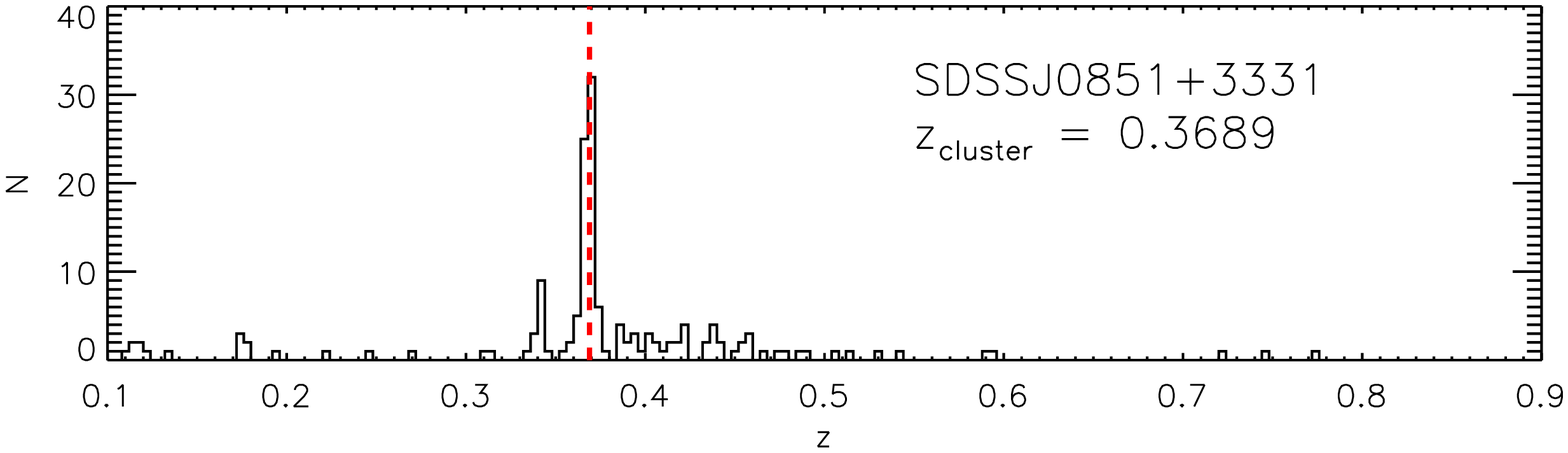}
\includegraphics[scale=0.35]{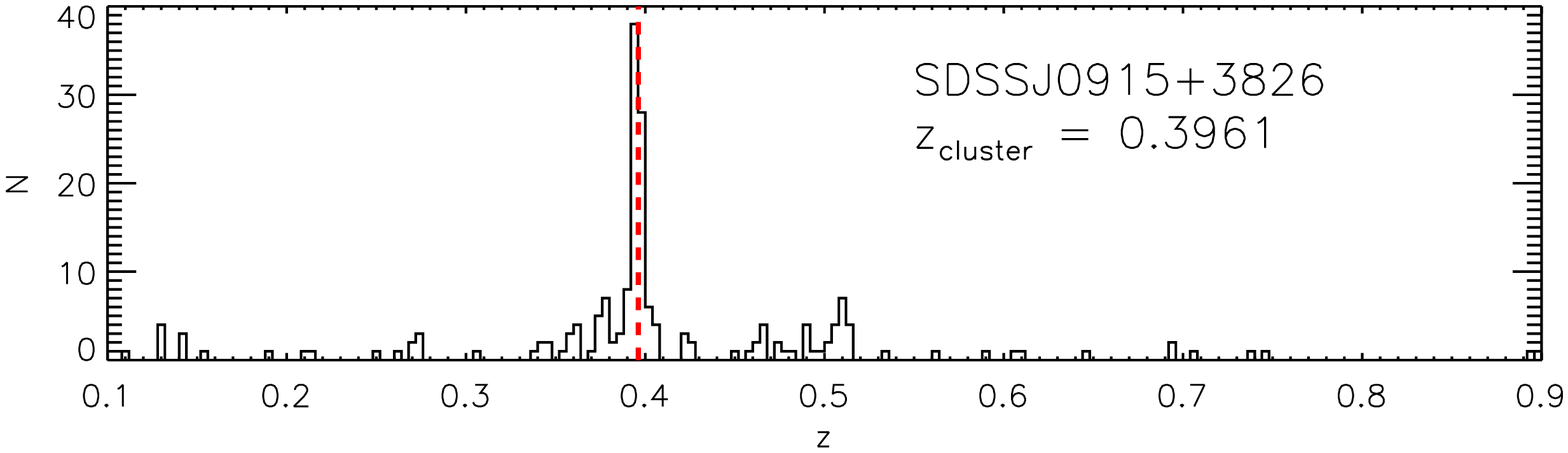}
\includegraphics[scale=0.35]{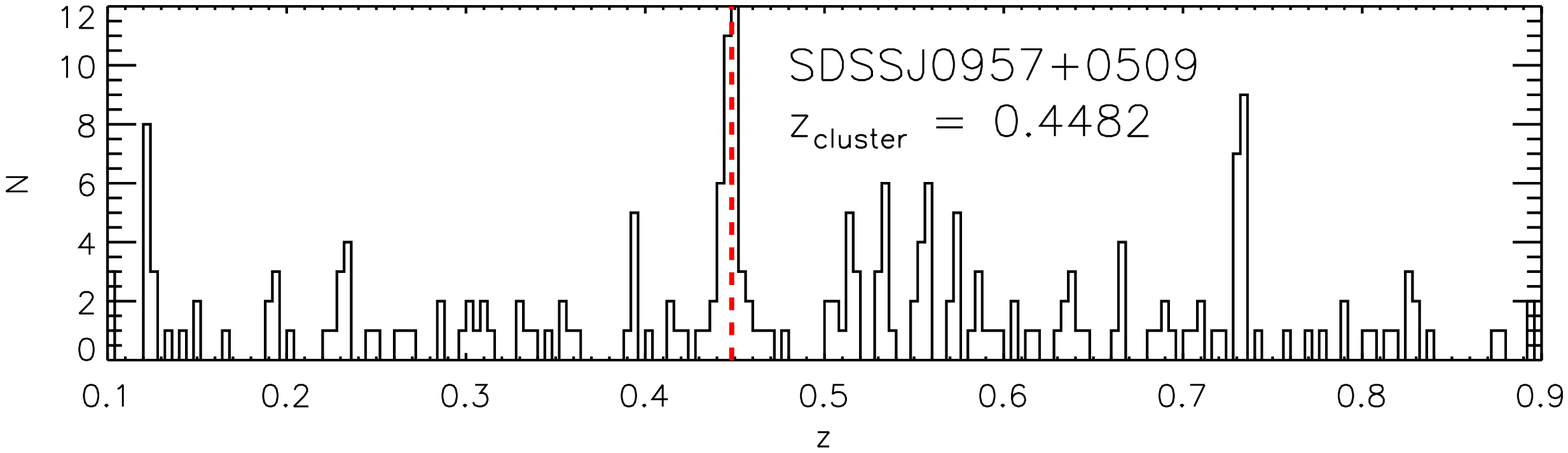}
\includegraphics[scale=0.35]{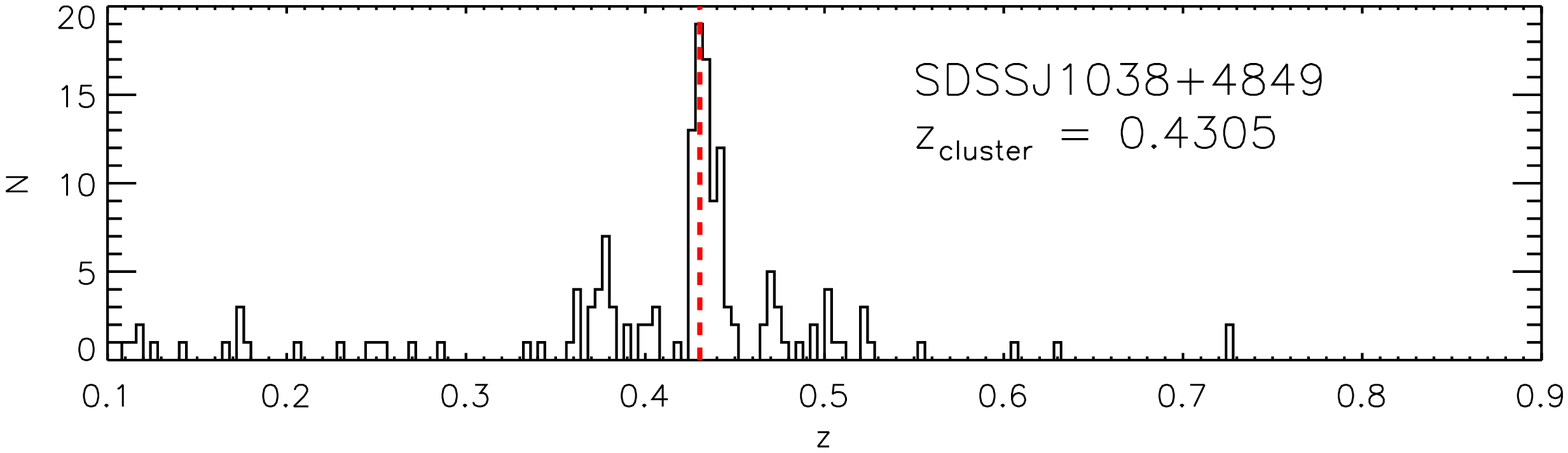}
\includegraphics[scale=0.35]{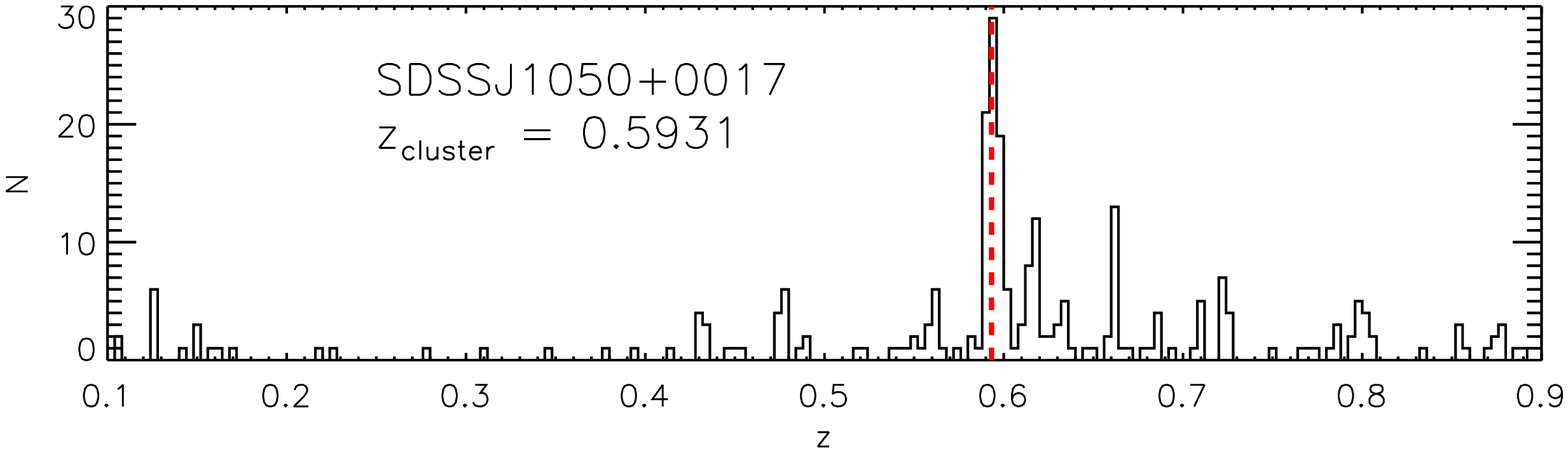}
\includegraphics[scale=0.35]{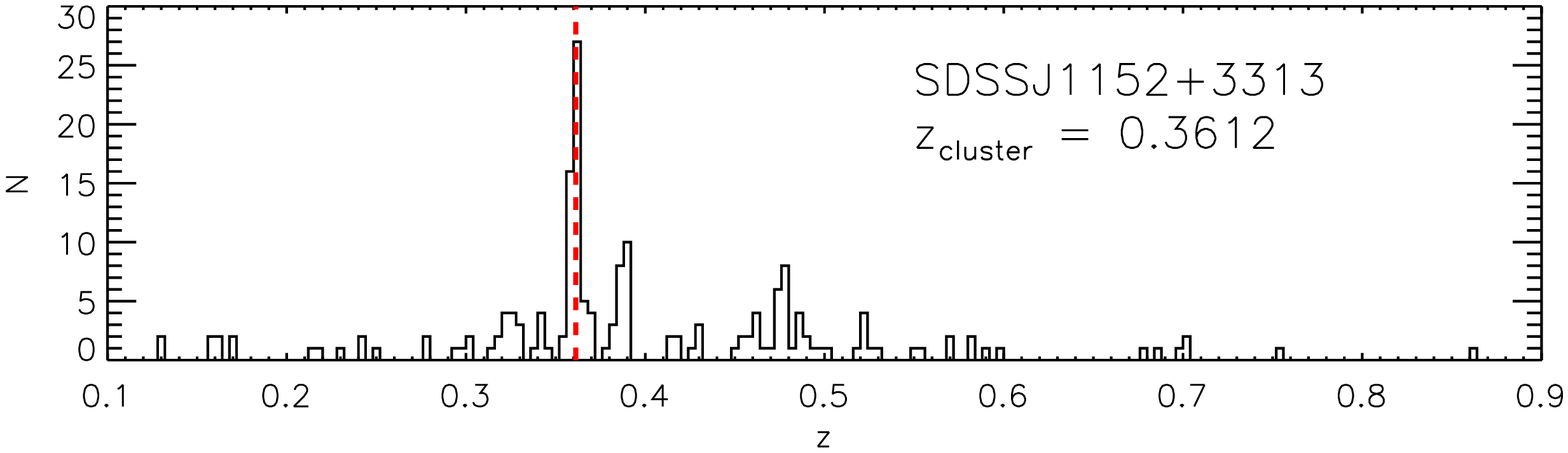}
\includegraphics[scale=0.35]{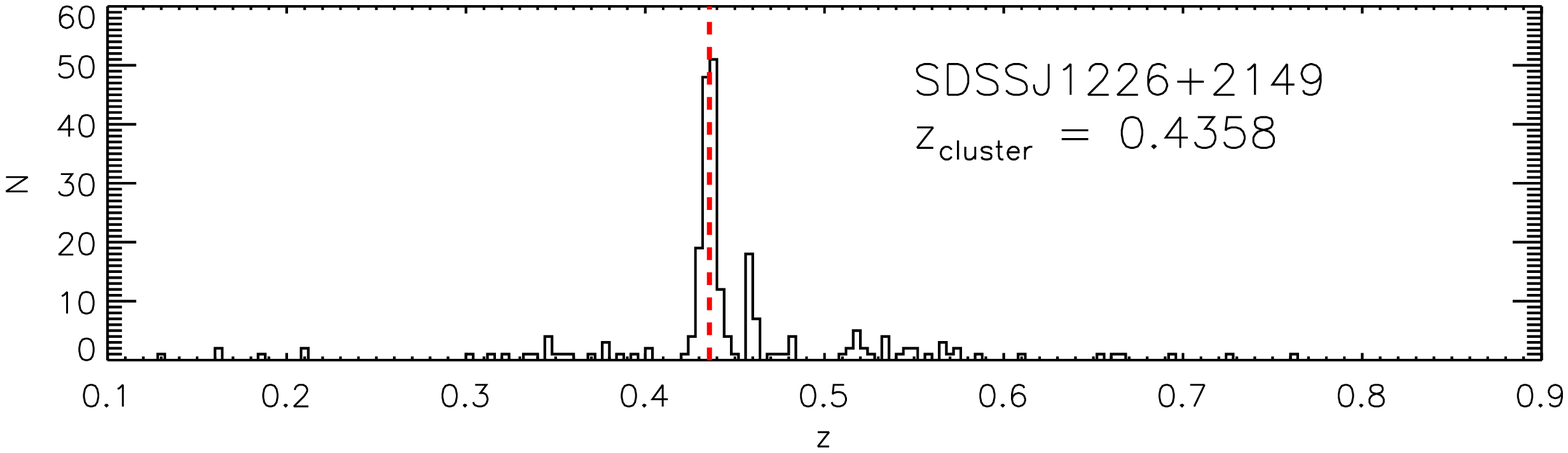}
\includegraphics[scale=0.35]{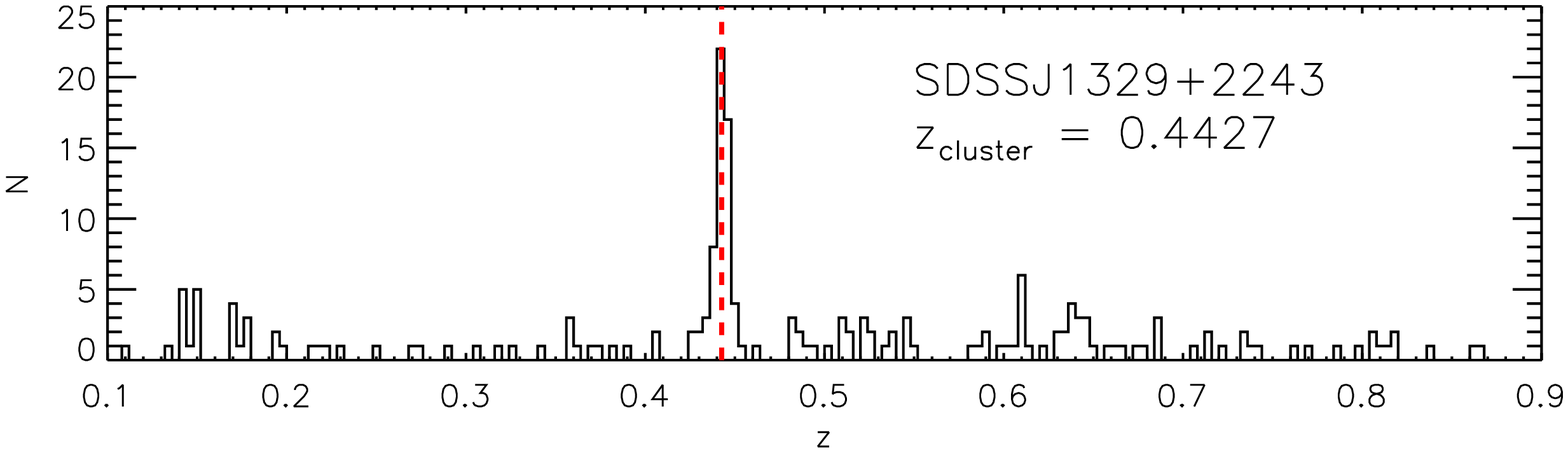}
\caption{\scriptsize{
Redshift distributions for the fields centered on the strong lensing clusters 
analyzed here, sorted into recession velocity bins of width 1200 \kms. The 
bi-weight median redshift of each strong lensing cluster is indicated by a 
vertical red dashed line. Note that apparent over-densities in the velocity 
space plotted here do not necessarily correspond to physical structures 
due to the spatial distribution of the individual redshifts on the sky.}}
\label{fig:zdist}
\end{figure*}



\subsection{MMT/Hectospec Spectroscopy}
\label{sec:hectospecobs}

The majority of the spectroscopy presented in this paper was obtained using the 
Hectospec instrument \citep{fabricant2005} at the 6.5m MMT Observatory on Mt. 
Hopkins, AZ. Hectospec places 300 fibers over a region in the sky approximately 
1 degree in diameter. In all fields observed with Hectospec fibers were placed on the 
sky using object selection and prioritization based on optical \emph{gri} colors from 
deep Subaru/SuprimeCam imaging \citep{oguri2012}. The total integration times for 
each field were 2-3$\times$1800~s. Hectospec data were reduced at the OIR 
Telescope Data Center and the Smithsonian Astrophysical Observatory using the 
pipeline of \citet{mink2007}, and redshifts measured using the RVSAO package 
\citep{kurtz1998}. Data were taken using the 270 line/mm grating, resulting in spectra 
with resolution, R $\sim$ 600--1500 (200-500 \kms), covering a wavelength range 
$\Delta \lambda =$ 3650-9200 angstroms. The RVSAO redshift uncertainties have 
been shown to be systematically underestimated by a factor of $\sim$2  
\citep[e.g.,][]{quintana2000}, so we use and report individual redshift 
measurements with uncertainties that are twice those that are output from RVSAO. 

\subsection{Magellan/IMACS Spectroscopy}
\label{sec:imacsobs}

Four clusters were observed with the Inamori-Magellan Areal Camera \& 
Spectrograph \citep[][]{imacsref} on the 6.5m Magellan-I (Baade) telescope at 
Las Campanas Observatory. SDSS J1226+2153 was observed using the f/4 
camera with the 
GISMO\footnote{www.lco.cl/telescopes-information/magellan/instruments/imacs/gismo/} 
module; GISMO re-images the central $\sim$3.5\arcmin~ region within the 
IMACS field of view so as to enable a factor of 8$\times$ spatial multiplexing of slit 
positions. A single GISMO mask places of order $\sim$100 slits within this region 
of the sky, allowing for dense spectroscopic sampling. Two GISMO slit-masks 
were observed on Apr 20 \& 21, 2009, one for 2$\times$1800~s in clear conditions 
and one for 2$\times$2400~s as clouds moved in. The second mask was re-observed 
in on May 27, 2009 due to the deteriorating cloud conditions limiting the quality of the 
data taken in Apr 2009; May observations were 3$\times$1800~s. The two masks 
were each centered on one of the two strong lensing cores in SDSS J1226+2153. 
The masks were observed with the 150 line/mm grating  and the WBP 4800-7800 
order-blocking filter, resulting in spectral resolution R $\simeq$ 450-700. The 
data were reduced using the COSMOS 
package\footnote{http://code.obs.carnegiescience.edu/cosmos}, along with 
custom IDL code. Custom IDL routines were also used to extract the spectra 
and measure redshifts. 

SDSS~J1050+0017 was observed using IMACS with the f/2 camera on UT Mar 17 
2013, and SDSS~J0957+0509 and SDSS~J1329+2243 were also observed with 
IMACS/f/2 on the UT Apr 09 2013. All of these observations used the 200 l/mm 
grism and the spectroscopic (i.e., no order blocking) filter and an unbinned 
detector, resulting in spectral resolution R $\simeq$ 500-1000 (300-460 \kms) 
and sensitivity over the wavelength range $\Delta\lambda =$ 4800-9800\AA. 
Two multi-slit masks were created for SDSS~J1050+0017, and each mask was 
observed for 3$\times$2400~s. A single mask was designed for each of 
SDSS~J0957+0509 and SDSS~J1329+2232, and these were exposed for 
3$\times$1500~s and 3$\times$1800~s, respectively. The IMACS spectra for 
SDSS~J1050+0017 are the same data used to inform strong lens modeling 
of that cluster lens in \citet{bayliss2013b}.

All masks were designed so as to place slits on a few faint, candidate strongly 
lensed background sources, with the remainder (and vast majority) of each mask 
devoted to placing slits on red-sequence selected cluster members and 
foreground/background field galaxies. All IMACS spectra were wavelength 
calibrated, bias subtracted, flat-fielded, and sky subtracted with the COSMOS 
reduction package, and then were extracted and stacked using custom IDL code.

\subsection{Gemini/GMOS-North Spectroscopy}
\label{sec:gmosobs}

SDSS J1050+0017 and SDSS~J1329+2243 also have previously unpublished 
redshift data from Gemini/GMOS-North that we include in our analysis. These 
data are almost identical to the spectra described in \citet{bayliss2011b}, with 
the exception that they were taken after the GMOS-North detectors were 
replaced with more sensitive e2vDD chips in November 2012. The primary goal 
of those observations was to obtain redshifts for lensed background sources, 
but open space in the mask was filled with red-sequence selected cluster 
members and other field galaxies.

The GMOS data were taken in macroscopic nod-and-shuffle mode, so that sky 
subtraction simply requires differencing two regions of the detector. The data 
were wavelength calibrated, extracted, stacked, flux normalized, and analyzed 
with a custom pipeline that uses the 
XIDL\footnote{http://www.ucolick.org/$\sim$xavier/IDL/index.html} package. The 
resulting spectra cover a wavelength range $\Delta \lambda =$ 5600-10000\AA, 
with a spectral resolution R $\simeq$ 700-1100 (270-430 \kms).  The data reduction 
and redshift measurements were nearly identical to that used by 
\citet{bayliss2011b}, with the only changes being updates made to account for 
the new e2vDD detectors.

\subsection{Supplemental Spectroscopy From the Literature}
\label{sec:literature}

We supplement the new spectroscopic measurements described above with 
redshifts from Gemini+GMOS North published in \citet{bayliss2011b}, SDSS 
DR10 \citep{sdssdr10} published redshifts, as well as redshifts for Abell 1703 
published by \citet{allen1992}, \citet{rizza2003}, and \citet{richard2009}. In total 
our redshift completeness as a function of $r$-band magnitude is consistently 
$\sim$10\% down to $r_{AB} =$ 22; this sparse sampling is not ideal, but still 
provides us with a rich dataset within which we can identify structures from the 
grouping of galaxies both in redshift and in spatial  distribution on the sky.

Example galaxies from our full redshift catalog are shown in 
Table~\ref{redshifttable}. The complete redshift catalogs used in this work have 
been made publicly available on the astronomy data repository on the Harvard 
Dataverse Network\footnote{http://thedata.harvard.edu/dvn/dataverses/cfa}.

\section{Redshift Sample and Identifying Structures}
\label{sec:redshift_analysis}

The redshift sample that we analyze here consists of all spectroscopic redshifts 
available from the data described in $\S$~\ref{sec:obs} that fall within the field 
of view of the SuprimeCam imager on the Subaru 8.2~m telescope (SuprimeCam 
photometry/astrometry was used for target selection and slit/fiber placement for 
the vast majority of the observations described above). This results in a spectroscopic 
sample covering a region on the sky with dimensions 
$\sim$34\arcmin$\times$27\arcmin. We analyze nine such fields in this paper that 
cover a total solid angle of 2.3 deg$^{2}$, containing ten unique strong lensing 
cluster cores.

\subsection{Spectroscopic Cluster Members}
\label{sec:members}

There is an obvious spike in the redshift distribution for each of our fields at the 
redshift of the primary strong lensing cluster (Figure~\ref{fig:zdist}). We select 
cluster members from the spectroscopic catalog beginning with a by-eye guess 
of the cluster redshift -- essentially the lens redshifts reported in \citet{oguri2012} 
-- and then compute an initial estimate using the bi-weight location and scale 
\citep{beers1990} of the velocity distribution for all redshifts within $\pm$0.02 in 
redshift, and within a projected physical radius, R$_{proj} \leq$ 1.5 Mpc. This 
projected radius corresponds approximately to the virial radius of these 
galaxy clusters. The choice of a physical cut in radius is somewhat arbitrary, 
but the resulting bi-weight median redshift estimates are insensitive to the 
exact choice. We then iteratively reject velocities separated from the 
bi-weight location by more than $\pm$ 3$\sigma$ until convergence is 
reached.The total number of redshifts, number of spectroscopically confirmed 
cluster members (selected via the procedure described above), and bi-weight 
median redshifts of each primary cluster lens are presented in Table~\ref{vtable}. 

\subsection{Identifying Line-of-Sight Structure}
\label{sec:overdensities}

%
%

\begin{figure*}
\centering
\includegraphics[scale=0.585]{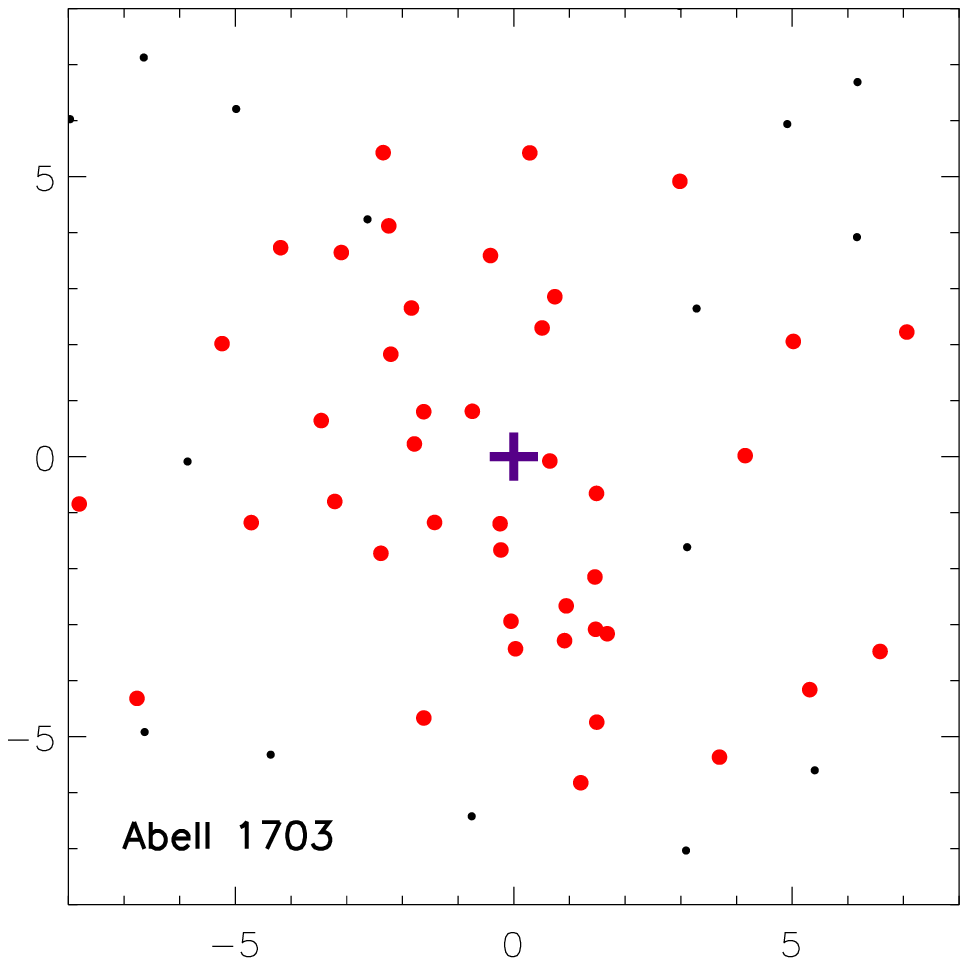}
\includegraphics[scale=0.585]{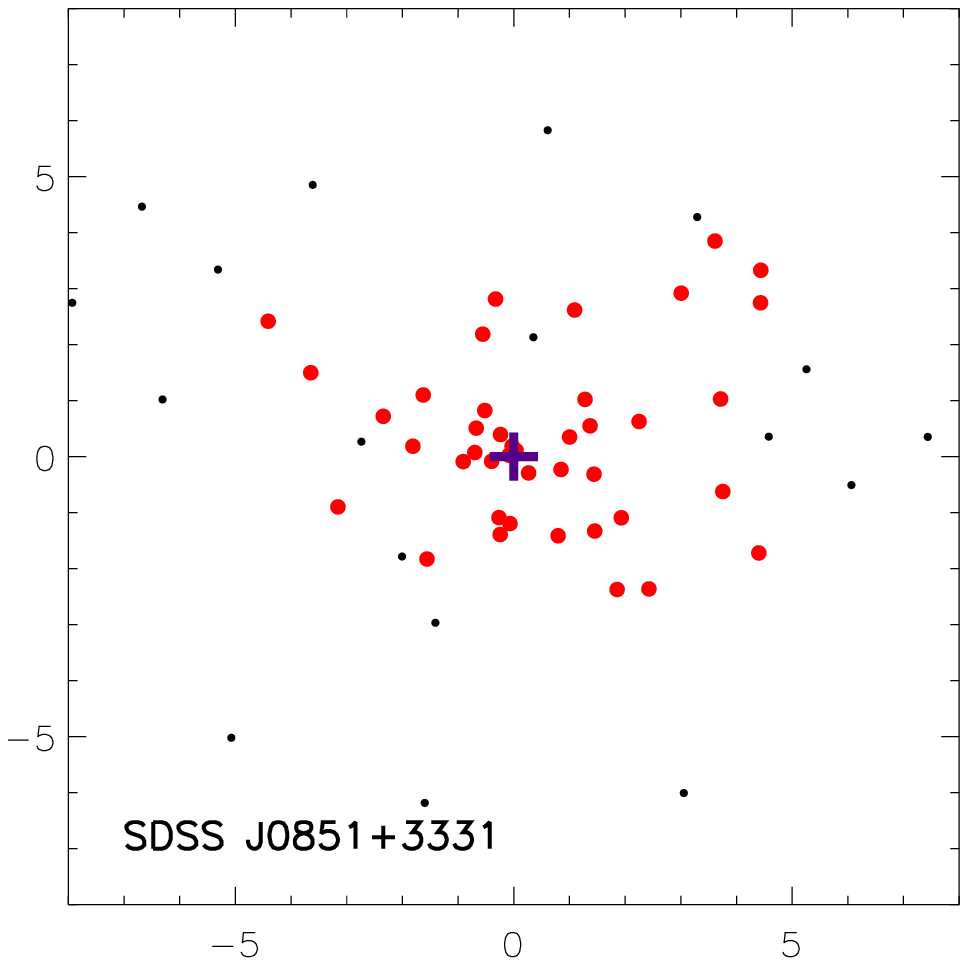}
\includegraphics[scale=0.585]{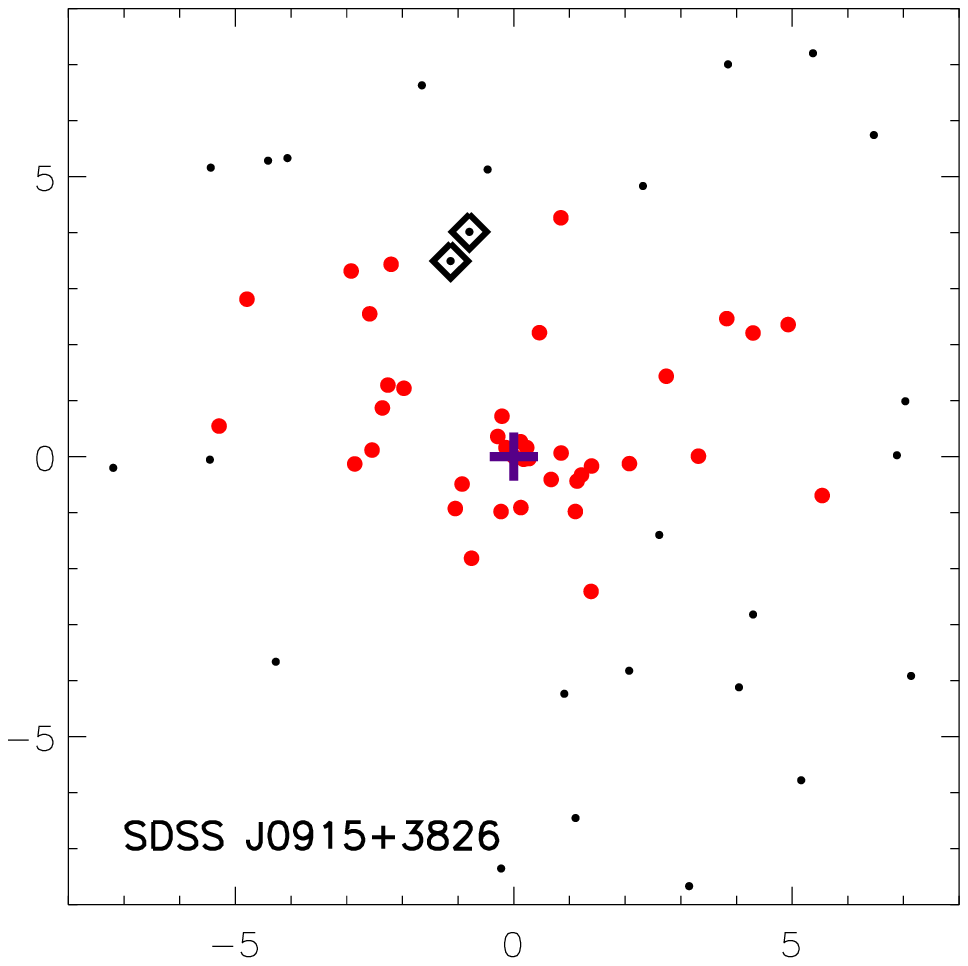}
\includegraphics[scale=0.585]{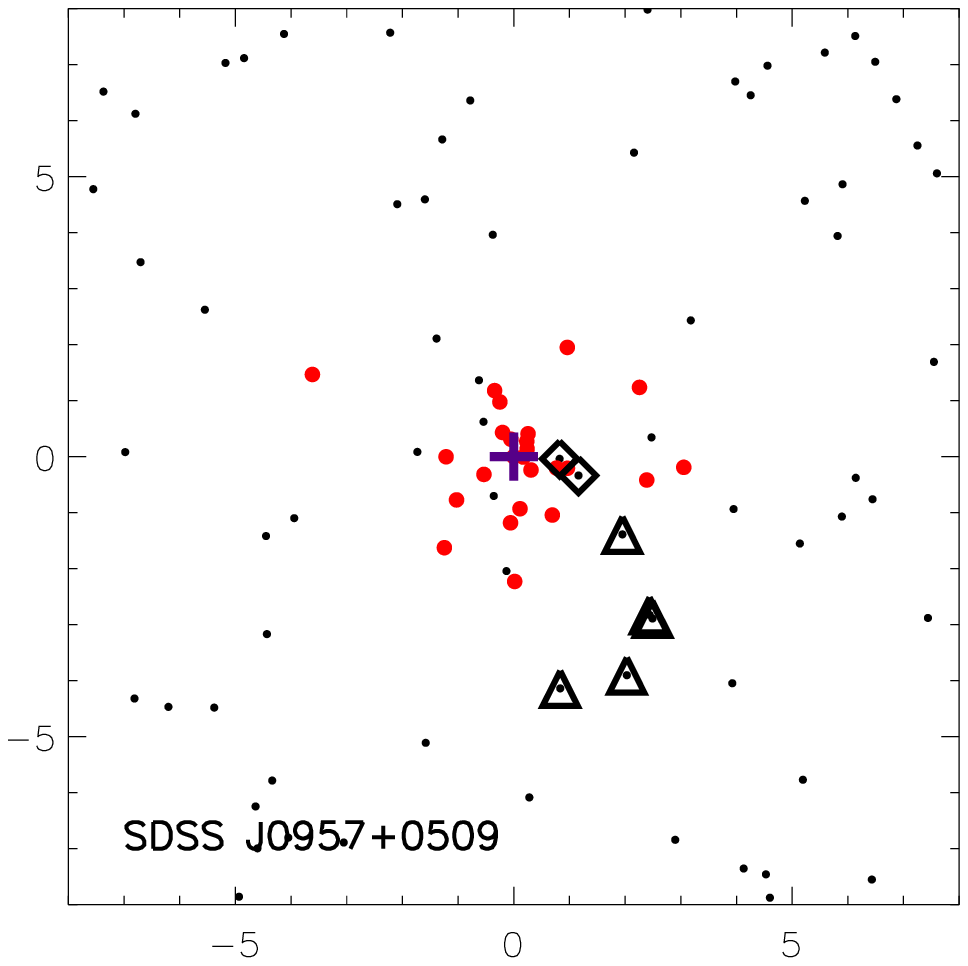}
\includegraphics[scale=0.585]{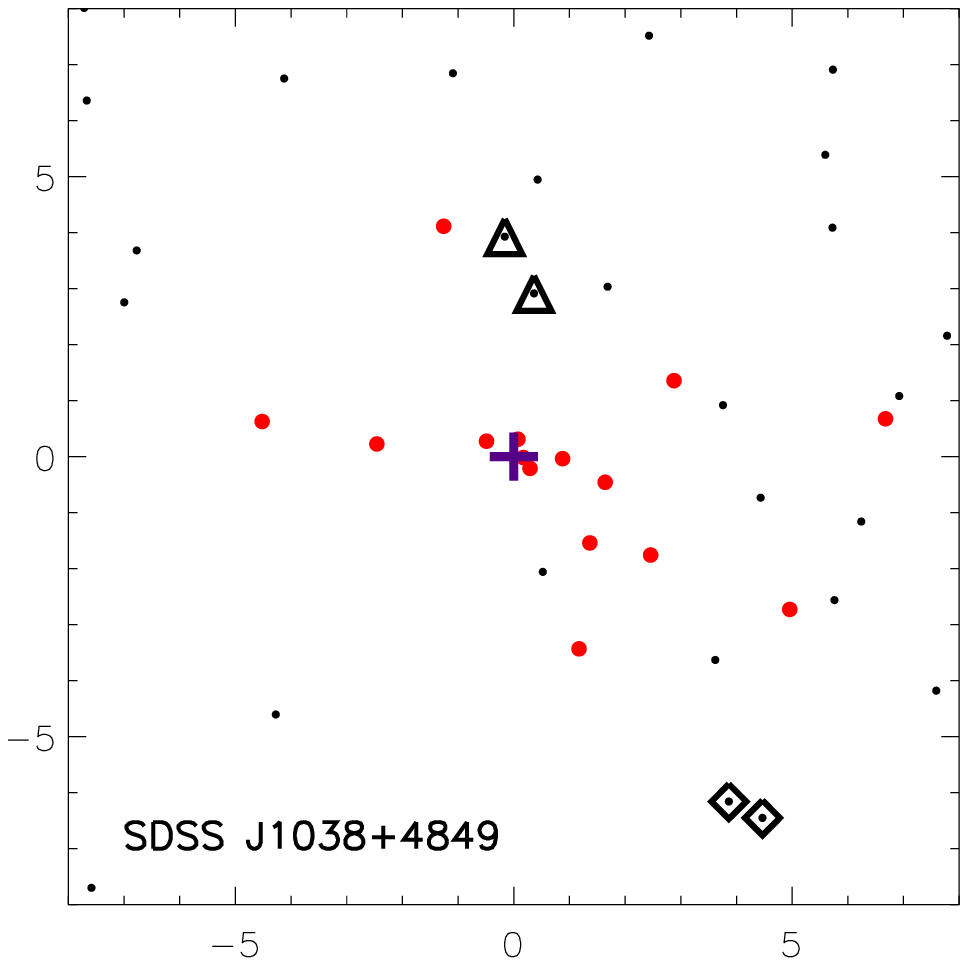}
\includegraphics[scale=0.585]{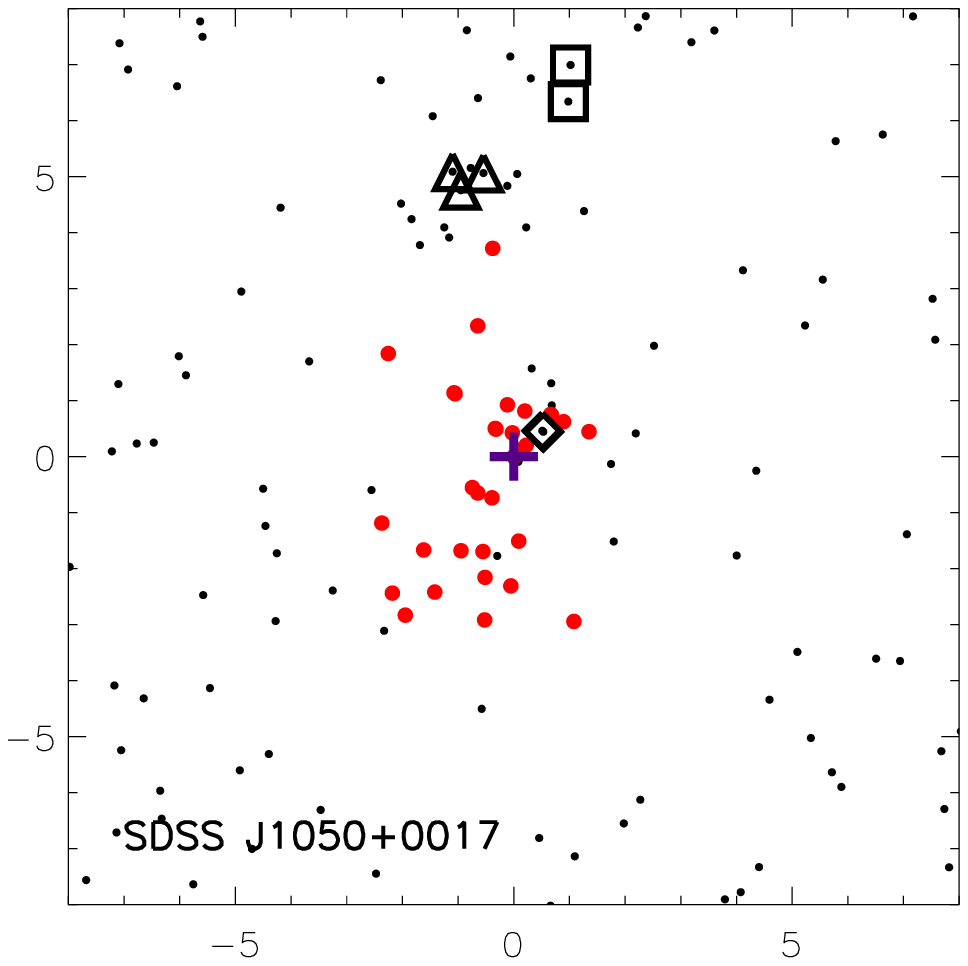}
\includegraphics[scale=0.585]{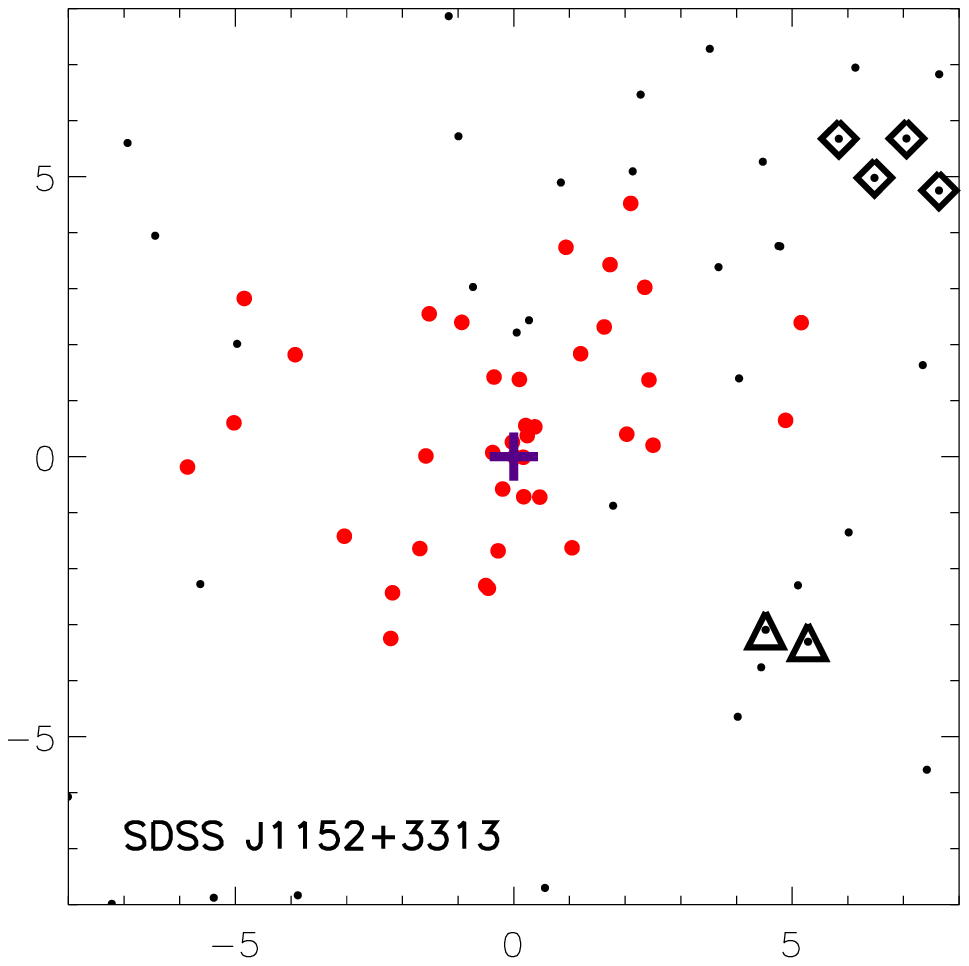}
\includegraphics[scale=0.585]{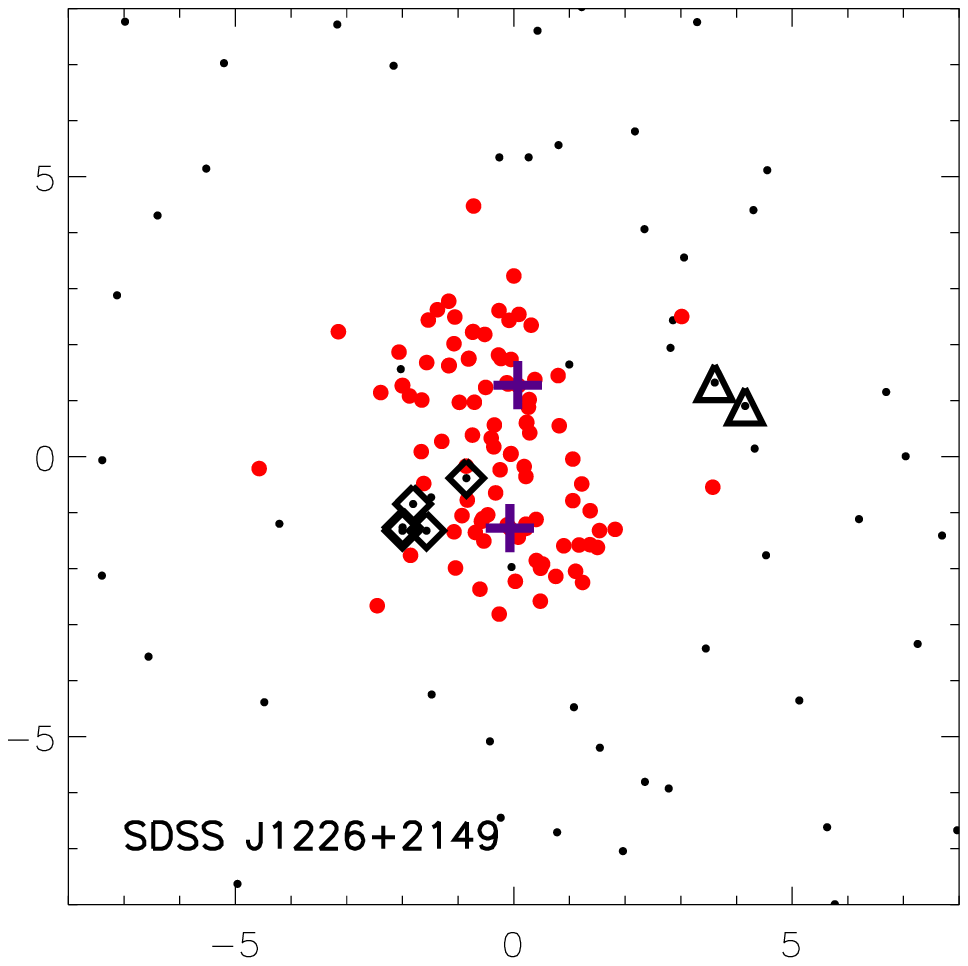}
\includegraphics[scale=0.585]{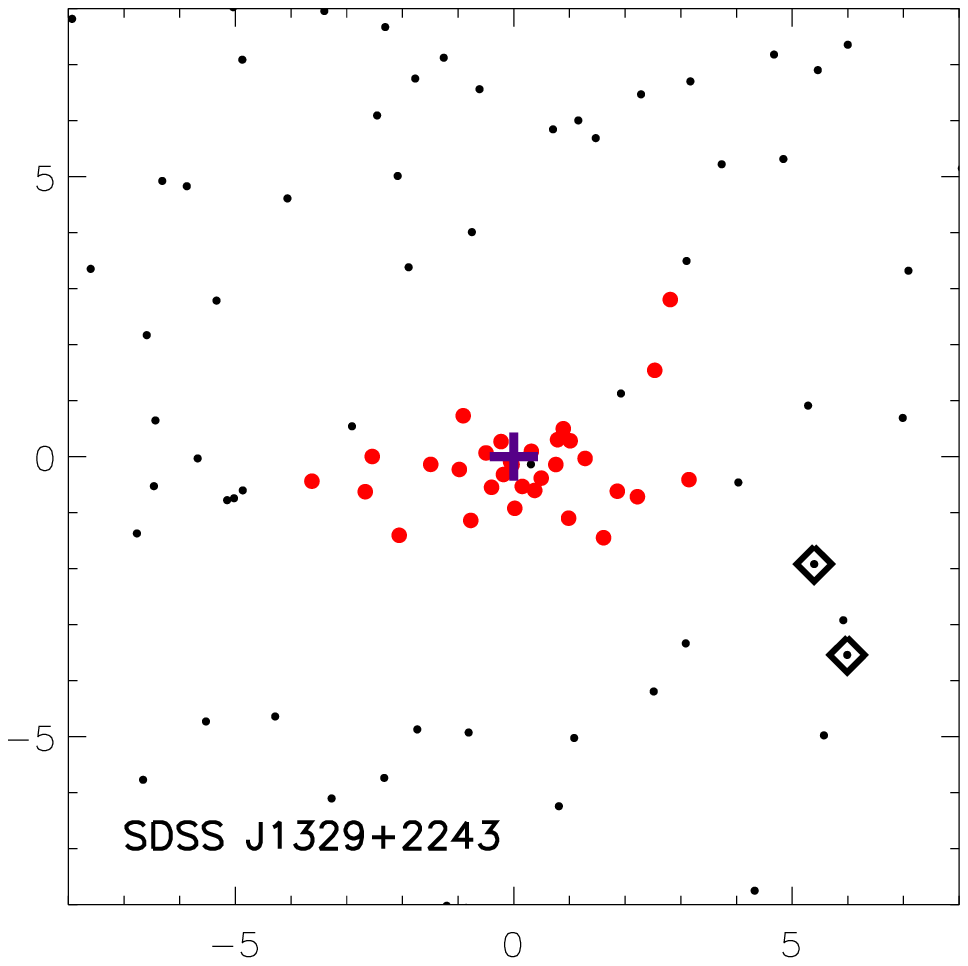}
\caption{\scriptsize{
The 16'$\times$16' region of the sky centered on each of the strong lensing cluster 
fields; the positions of all galaxies with spectroscopic redshifts are plotted using different 
symbols to indicate type. Filled larger red dots are cluster member galaxies, filled smaller 
black dots are field galaxies that are not associated with the cluster. Galaxies that are 
associated with one of the groups identified in $\S$~\ref{sec:overdensities} are also 
indicated by an open black plot symbol, with different symbols corresponding to different 
groups identified in a given field. The centroid of the strong lensing is always indicated by 
a blue cross (SDSS~J1226 has two cluster cores that are strong lenses, each is 
indicated by its own blue cross). In the line of sight centered on SDSSJ~1050+0017 
there is a group located near the lensing cluster core with N $=$ 2 spectroscopic 
members that are separated by $\sim$2\arcsec.
}}
\label{fig:zsky}
\end{figure*}

We identify additional structures by looking for over-densities in distribution of 
spectroscopically measured galaxies along the line of sight toward each cluster 
lens. We use a prescription based on the group catalog that was defined in the 
zCOSMOS 10k redshift survey \citep{knobel2009}. \citet{knobel2009} describe 
two methods for identifying group structures from redshift catalogs, and evaluate 
the performance of these methods using mock catalogs from simulations. Here 
we adopt the ``Friends of Friends'' (FoF) method, as it is the simpler of the two, 
having fewer free parameters than the alternative (VDM) algorithm, and the 
performance of the two methods is extremely similar.

\begin{deluxetable}{lcccc}
\tablecaption{Line-of-Sight Velocity Data\label{vtable}}
\tablewidth{0pt}
\tabletypesize{\tiny}
\tablehead{
\colhead{Cluster Name} &
\colhead{ N$^{a}_{z}$ } & 
\colhead{ N$^{b}_{c}$ } &
\colhead{ z$_{cluster}$} &
\colhead{ $\sigma_{v}$ (km s$^{-1}$)}  
}
\startdata
Abell1703  &  182 &  42 &  0.2770 $\pm$ 0.0010  & 1380 $\pm$ 140  \\
SDSSJ0851+3331  & 169 &  41 &  0.3689 $\pm$ 0.0007  & 890 $\pm$ 130 \\
SDSSJ0915+3826  &  218 &  39 &  0.3961 $\pm$ 0.0008  & 960 $\pm$ 120  \\
SDSSJ0957+0509  &  280  &  25  & 0.4482 $\pm$ 0.0010 & 1250 $\pm$ 290  \\
SDSSJ1038+4849  &  168 &  15 &  0.4305 $\pm$ 0.0008  & 550 $\pm$ 90   \\
SDSSJ1050+0017  &  499 &  32 &  0.5931 $\pm$ 0.0005  & 560 $\pm$ 80  \\
SDSSJ1152+3313  & 204 &  38 &  0.3612 $\pm$ 0.0007 & 800 $\pm$ 90   \\
SDSSJ1226+2149\tablenotemark{c}  &  252 & 98 &  0.4358 $\pm$ 0.0004  & 870 $\pm$ 60  \\
SDSSJ1329+2243  &  248  &  31  & 0.4427 $\pm$ 0.0007   &   830 $\pm$ 120
\enddata
\tablenotetext{a}{~Size of the total spectroscopic galaxy sample in each field.}
\tablenotetext{b}{~Number of cluster members within a projected physical 
radius of 1.5 Mpc.}
\tablenotetext{c}{~This is a complex system that includes three different 
cluster-scale structures separated in recession velocity by a few hundred 
\kms~; two of the three are strong lenses and separated by 
$\sim$2.5\arcmin~on the sky.}
\end{deluxetable}

The parameters of the FoF group finder are tuned to detect groups with 
different numbers of members (ie., N $\geq$ 6, N $=$ 5, N $=$ 4, N $=$ 3, 
and N $=$ 2), based on extensive testing in mock catalogs \citep{knobel2009}. 
We apply the FoF finder with these same optimized parameters (along with the 
galaxy number density computed from our own data) after removing the member 
galaxies of the primary strong lensing clusters-- where member galaxies are 
defined from the criteria described in $\S$~\ref{sec:members}. We characterize 
each group with a position on the sky equal to the mean right ascension and 
declination of the identified group members, and with a redshift computed using 
the bi-weight location estimator. From these quantities we measure the 
projected angular distance on the sky and the comoving distance along the 
line of sight (i.e., perpendicular to the plane of the sky) between each strong 
lensing cluster core and the groups identified along their lines of sight. Example 
groups identified in our data are shown in Table~\ref{tab:exgroups}. Where 
there are sufficient members to measure a dispersion \citep[N $\geq$ 4 using the 
gapper method;][]{beers1990} we find that the over-densities identified in our 
data have very small dispersions ($\sigma_{v} \lesssim$ 300 \kms; e.g.,
Table~\ref{tab:exgroups}).

Structure in the universe is strongly correlated 
\citep[e.g.,][]{mo1996,tadros1998}, and the more massive the structure the 
stronger the correlation \citep{bahcall2003,estrada2009}. Structures that are 
separated by sufficiently small comoving radial distances are significantly 
more correlated than those with large comoving separations. Structure is, of 
course, correlated at a non-zero level out to extremely large scales (i.e. 
hundreds of Mpc), but as a practical matter here we seek to identify 
``uncorrelated'' structures as those structures that are not associated or 
interacting/merging with the primary lensing cluster. With a well-defined 
group sample in-hand, we now consider how to differentiate between groups 
that trace uncorrelated line-of-sight structure and groups that are located close 
enough to the primarily cluster lenses to be potentially interacting with them. 
It has also been shown that the mass distributions of massive clusters are 
aligned with surrounding filamentary structure, both in simulations \citep{noh2011} 
and in the SDSS \citep{smargon2012}. These alignments are a manifestation 
of exactly the sort of correlated line of sight structure that we want to 
remove.

Depending on how aggressively we want to define ``uncorrelated'' 
we can make several cuts on our group sample based on the comoving 
separation between each group and their associated primary cluster lenses. 
\citet{bahcall2003} measured the correlation length to grow from $\sim$16 
Mpc to $\sim$34 Mpc for groups and rich clusters, respectively; these 
numbers provide guidance for what cuts on the comoving distance we 
should use to pick out the structure that is uncorrelated. For all analyses 
that follow we make two such cuts: 1) the more aggressive cut 
labels uncorrelated structures as only groups that are separated by a comoving 
distance, D$_{m}$ $>$ 100 Mpc from the primary cluster lenses, and 2) the 
less aggressive cut uses D$_{m}$ $>$ 30 Mpc. 

\begin{figure}[h]
\centering
\includegraphics[scale=0.43]{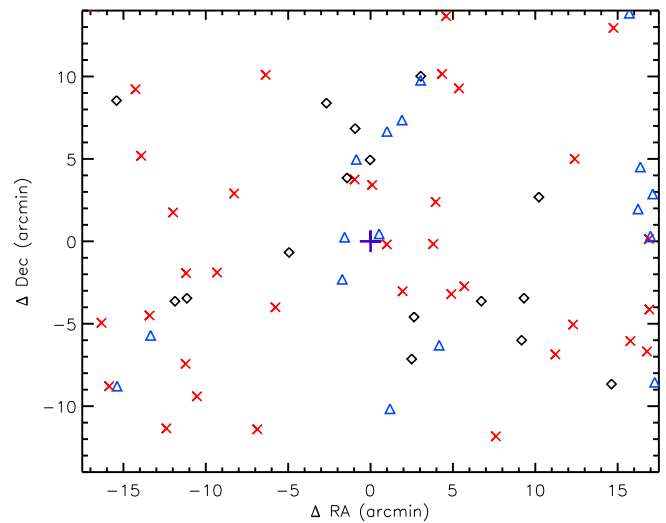}
\caption{\scriptsize{
All groups identified in our spectroscopic data are plotted at their 
positions relative to the 10 strong lensing cluster cores analyzed in this work 
(SDSS J1226+2149 has two distinct strong lensing cores). The fiducial strong 
lensing centers are identified with the purple cross, and groups are split into 
two groups. Red X's indicate groups that are separated from the primary lensing 
clusters by $>$ \cmhigh~Mpc along the line of sight; these are structures that 
we can confidently label as uncorrelated with the primary cluster lens. Blue 
triangles indicate groups that are separated by $>$ \comdlimit~Mpc but less 
than \cmhigh~Mpc  -- these are also likely to be uncorrelated with the primary 
cluster lenses. Black 
diamonds indicate groups separated by $\leq$ \comdlimit~Mpc along the line 
of sight, and are therefore very possibly correlated with the primary cluster lens 
(we do not consider these groups in our measurement of the uncorrelated 
line-of-sight structure). The plot displays a 35\arcmin$\times$28\arcmin~field of 
view, which matches the field covered by the Subaru/SuprimeCam imaging that 
was used for fiber/slit placement in our spectroscopic observations.
}}
\label{fig:allgroups}
\end{figure}

In Figure~\ref{fig:zsky} we plot the spectroscopic data -- with all groups separated 
by a comoving distance of at least 30 Mpc and cluster members marked -- in the 
regions of sky centered on each primary lensing cluster. Figure~\ref{fig:allgroups} 
shows the distribution of group positions relative to their associated strong lensing 
clusters, with both the 30 Mpc and 100 Mpc cuts indicated. The final group catalog 
along each strong lensing selected line of sight is also summarized in 
Table~\ref{tab:groupsummary}.

\begin{deluxetable}{cccccc}
\tablecaption{Example Groups\label{tab:groupcat}}
\tablewidth{0pt}
\tabletypesize{\tiny}
\tablehead{
\colhead{$\alpha$ } & 
\colhead{ $\delta$ } &
\colhead{ z } &
\colhead{ N$_{gal}$ } &
\colhead{ $\sigma_{v}$\tablenotemark{a} } &
\colhead{ $\theta_{proj}$ }  \\
\colhead{ J2000 } & 
\colhead{ J2000 } &
\colhead{~~ } &
\colhead{~~ } &
\colhead{ (\kms)} &
\colhead{ (arcmin) } 
}
\startdata
13 14 52.2 & +52 03 35 & 0.30280 &  2 & --- &   14.7  \\
13 15 46.1 & +51 51 44 & 0.27566 &  2 &  --- &    6.9  \\
13 16 14.2 & +51 40 30 & 0.29869 &  4 &  195 &   13.7  \\
13 13 26.4 & +51 49 44 & 0.38489 &  2 & --- &   15.3  \\
13 14 39.8 & +51 59 08 & 0.05955 &  5 &  141 &   10.8  \\
13 14 20.3 & +51 41 37 & 0.10094 &  2 &  --- &   10.2  \\
13 15 23.6 & +52 02 42 & 0.05971 &  2 &  --- &   13.9 
 \enddata
\tablenotetext{a}{~Computed using the gapper statistic \citep{beers1990} only for 
groups with N$_{gal}$ $\geq$ 4.}
\label{tab:exgroups}
\end{deluxetable}

\subsection{Selection Effects In the Galaxy Redshift Catalogs}
\label{sec:selectioneffects}

The catalog of spectroscopic redshifts that we have for each cluster is subject to 
important observational selection effects that we must take into account before 
interpreting the results of the preceding analysis. Most importantly, the multi-slit 
and multi-fiber spectroscopic observations that we use were designed with the 
primary goal of measuring cluster member redshifts in each primary lensing 
cluster. We used a red sequence and blue cloud selection at the lensing cluster 
redshift to give higher weights/priorities to the placement of slits/fibers onto likely 
cluster members. In practice, only a modest fraction of slits/fibers could be placed 
on likely members in any given slit mask or fiber configuration (due to practical 
considerations such as limited source density of candidate members and 
slit/fiber collisions).

From Table~\ref{vtable} we see that cluster members always make up less than 
45\% of our redshift sample, and typically they represent only $\sim$15-20\% of 
redshifts in a given cluster lens field. Thus, while we recover many non-cluster 
member redshifts in each of our fields, our observational strategy nevertheless 
biases us against identifying line-of-sight structures, in favor of better sampling 
the primary lensing cluster velocity distributions. 

We also note that the cluster members are typically spatially concentrated in 
the core of each primary cluster lens, so that our non-cluster member redshifts 
are preferentially at larger angular separations from the lensing cluster than 
they would have been had the slit/fiber placement had been purely random (or 
based purely on a simple criterion such as magnitude). This effectively biases 
us against identifying projected structures with small projected angular separations 
from the lensing clusters. This bias is additive with the one discussed above in 
that it hinders our ability to identify groups in our spectroscopic data. The bias 
against identifying line-of-sight structures near the cores of our cluster lenses 
is likely most evident in the lines of sight centered around Abell 1703 and 
SDSS~J0851+3331; the spectroscopic catalog for both of these lines of sight is 
dominated by cluster member galaxies within a 10\arcmin~radius of the strong 
lensing cluster cores, and we identify no FoF groups within these regions.

Our method of identifying candidate line-of-sight structures depends solely on the 
locations of galaxies in the sky and in recession velocity. It is, of course, also 
possible to search for group and cluster-like structures using a red sequence 
selection in the available imaging. However, such a selection would only find 
evolved/collapsed structures, while possibly missing small group-like structures 
that are still in the process of forming and have not yet developed a 
population of passively evolving member galaxies. Additionally, the purity and 
completeness of a red sequence selection becomes significantly worse in the 
smallest structures (groups with very low numbers of galaxies) 
\citep[e.g.,][]{koester2007}. These smaller, lower-mass -- and often less-evolved -- 
structures are far more numerous than more massive groups and are therefore 
most likely to produce a chance line-of-sight alignment with a massive cluster 
lens. Because these chance alignments are precisely what we wish to measure, 
a simple red-sequence search for line-of-sight structure is not ideal for the 
analysis that we perform in this paper.

\subsection{Quantifying of Line-of-Sight Structure}

We want to test the hypothesis that strong lensing clusters preferentially lie 
along lines of sight that have systematically more intervening structure than 
would be expected along random lines of sight. Quantifying the excess (or 
dearth) of structure along the line of sight toward our strong lensing clusters 
therefore must be compared against a measurement of the amount of 
intervening structure along random lines of sight.  
We previously adopted the group-finding algorithm that was used to produce 
the zCOSMOS 10k group catalog, making it an ideal dataset for comparing 
against our measurements.The zCOSMOS 10k group catalog is defined from 
a spectroscopic catalog that includes $\sim$6500 redshifts in the interval, 0.1 
$<$ z $<$ 0.8, from a 1.7 deg$^{2}$ region on the sky \citep{knobel2009}. In 
comparison, our data include 1707 redshifts in the same interval, covering 
a 2.3 deg$^{2}$ area on the sky. Our coverage is therefore $\lesssim$25\% 
as dense as the zCOSMOS 10k catalog, so that any comparison of groups 
identified in our data will necessarily represent a \emph{lower limit} on the 
number of groups that would be identified in a dataset with spectroscopic 
coverage as dense as the zCOSMOS 10k catalog. 

Because our spectroscopic data is more sparsely sampled that the zCOSMOS 
10k catalog it is not valuable to compare the simplest observable 
quantities, such as a count of the number of groups that are projected within 
various angular separations of our strong lensing clusters. We can, however, 
measure the distribution of angular separations between our strong lensing 
clusters and the groups that we identify along each line of sight. The 
sparseness of our spectroscopic data should repress the total number of 
groups found, but it should not impact the spatial distribution of those groups. 
An over-abundance of structure along the line of sight toward our 
strong lensing cluster sample should produce a signal in the 
angular separations between the cluster lenses and the groups that we 
identified.

We measure the distribution of angular separations between strong lensing 
clusters and uncorrelated groups as follows. For each strong lensing cluster 
core we select groups that fall within a 20\arcmin~radius of the centroid of the 
strong lensing (i.e., the center of the mass distribution of the primary cluster 
lens) and measure the radial angular separations between strong lensing 
core and groups for each core-group pair. The 20\arcmin~cutoff represents 
the maximum radius out to which our redshift and group catalogs extend 
around all of our strong lenses. All of these measurements are made using 
the group catalogs with both D$_{M}$ $>$ 30 Mpc and D$_{M}$ $>$ 100 
Mpc cuts to isolate uncorrelated groups (see $\S$~\ref{sec:overdensities}). 

\begin{figure}
\centering
\includegraphics[scale=0.43]{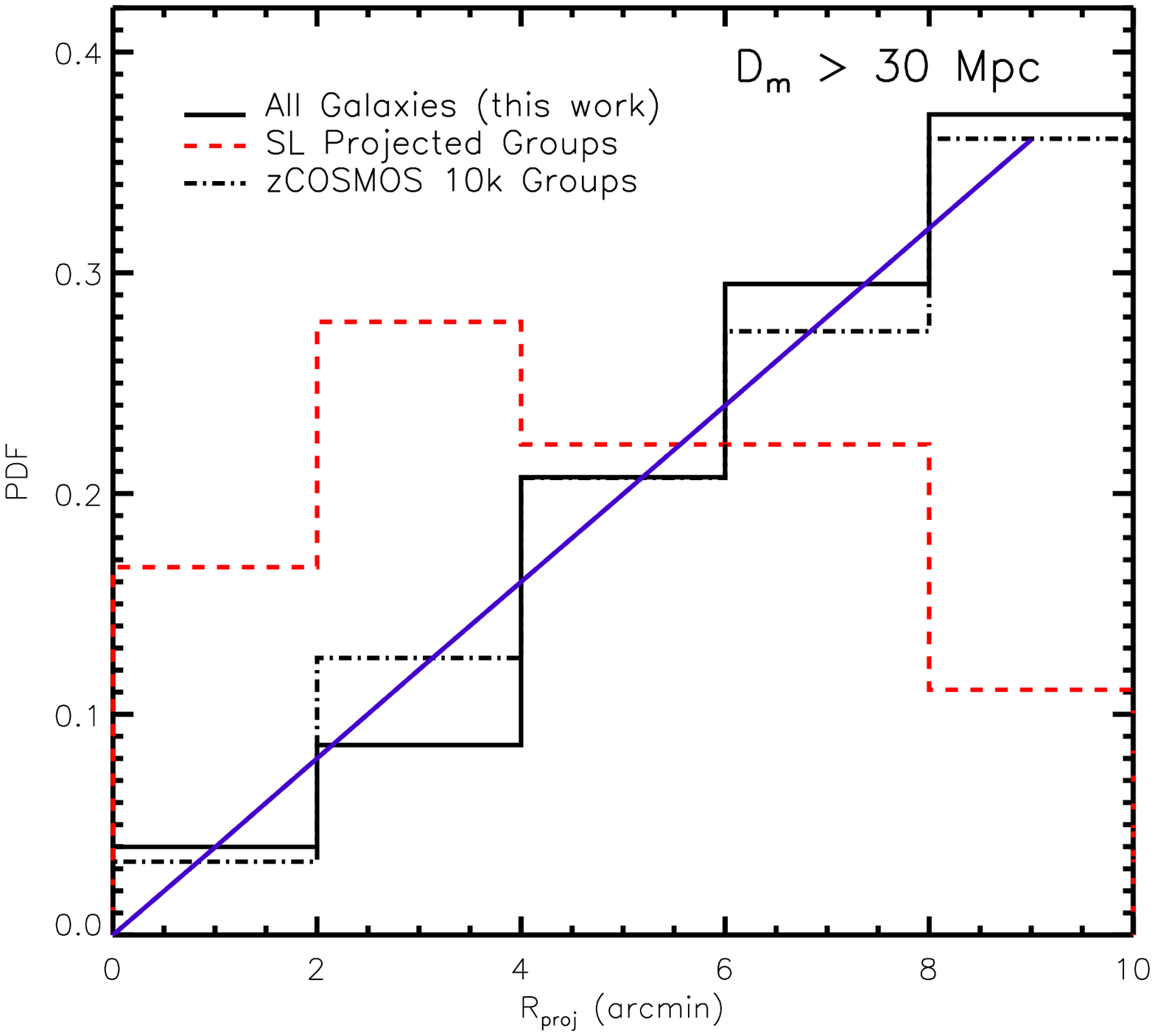}
\includegraphics[scale=0.43]{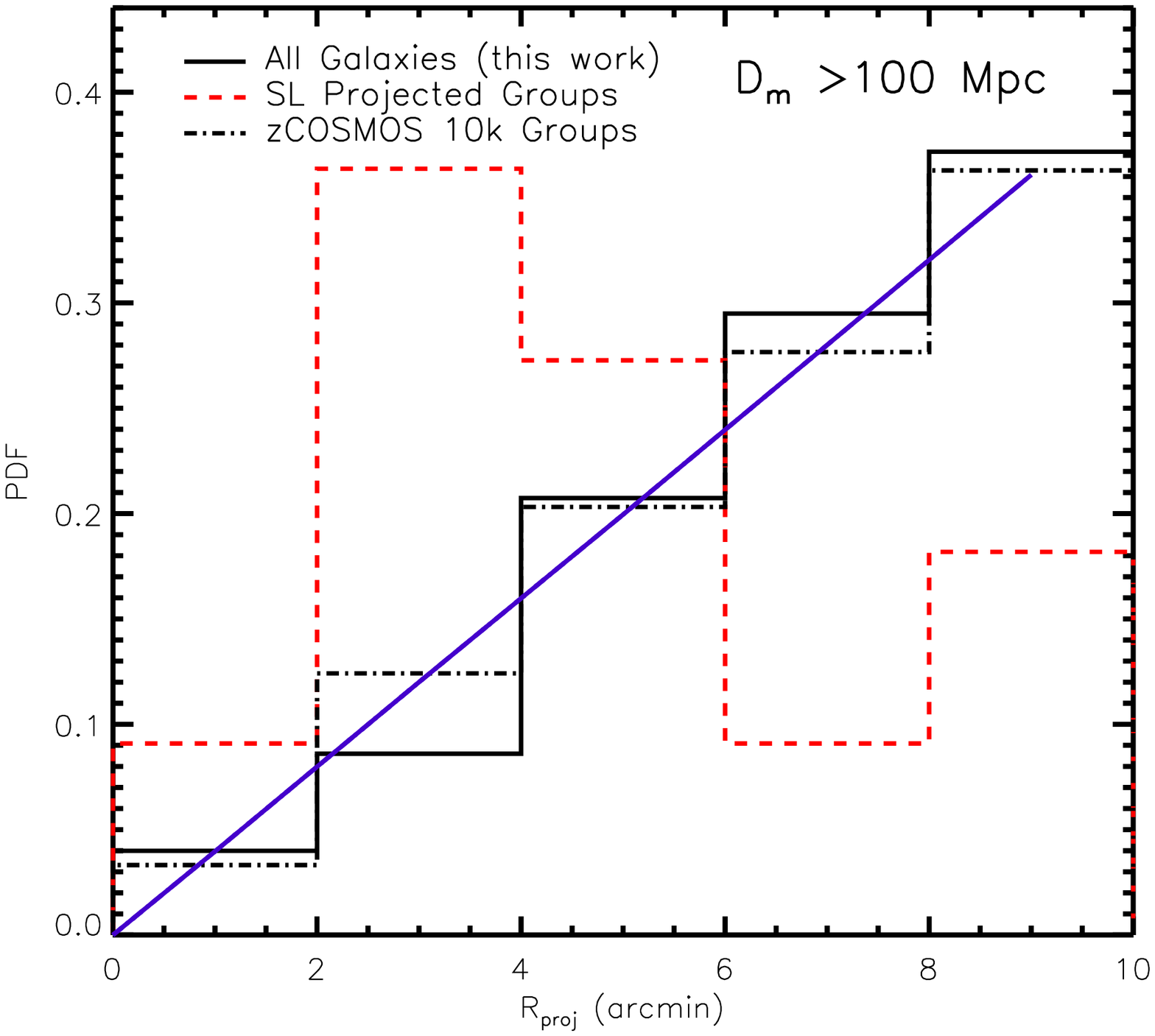}
\caption{\scriptsize{
\emph{Top:} 
The distribution of angular separations from the centroid of the strong lensing for both 
1) all non-cluster member galaxies with spectroscopic redshifts (solid line), and 2) the 
groups identified in $\S$~\ref{sec:overdensities} using the friends of friends algorithm 
(dashed red line). For comparison we also show the distribution of radial separations 
between large groups ($\sigma_{v} \geq$ 500 \kms) and other uncorrelated structures 
in the zCOSMOS 10k group catalog (dot-dashed line). The plotted distribution is the 
probability distribution function of angular separations from the strong lensing 
cluster cores (or the zCOSMOS groups with $\sigma_{v} \geq$ 500 \kms) extending 
out to a radius of 20\arcmin. We use a coming separation of 30 Mpc 
from the primary cluster lenses as the cut to define uncorrelated structure. We also 
plot the expected distribution of projected angular separations around a fiducial 
cluster along a random line of sight, as predicted by the 2 point correlation function 
and after also removing structures within a comoving distance of 30 Mpc (purple 
solid line).
\emph{Bottom:}  Same as the top panel, but with a cut of 100 Mpc from the 
primary cluster lenses as the cut to define uncorrelated structure.}}
\label{fig:rdist}
\end{figure}

For comparison against our strong lensing selected cluster sample we also 
measure the radial separations in the same way for our redshift catalog (the 
input for the FoF group finder), and also for the zCOSMOS 10k group catalog 
where we we use the most massive groups in the catalog ($\sigma_{v} \geq$ 
500 \kms) as the centroids for measuring radial separations. 
In Figure~\ref{fig:rdist} we plot the probability distribution functions of angular 
separations between each of: 1) strong lensing clusters and uncorrelated 
groups, 2) strong lensing clusters and all redshift data, and 3) zCOSMOS 
massive groups and other uncorrelated groups in the zCOSMOS catalog.
Figure~\ref{fig:rdist} also shows the expectation for the observed angular 
separation distribution along a random line of sight that is predicted from 
the 2-point correlation function, which has been measured for group and 
cluster scale structures \citep[as mentioned previously in 
$\S$~\ref{sec:overdensities};][]{bahcall2003}. We generate this prediction 
by populating a cosmological volume with structures around a fiducial cluster 
at z $=$ 0.43 (the median redshift of our cluster lens sample), where these 
structures are drawn by Monte Carlo from the 2 point correlation function 
with correlation lengths of r$_{0}$ = 10, 20, and 40 Mpc. We then apply 
the same cuts (D$_{m}$ $>$ 30 \& 100 Mpc) to remove simulated structures 
that are nearby the fiducial cluster and then measure the resulting 
distribution of projected angular separations. The resulting distributions 
are insensitive to the choice of correlation length, as should be expected 
given that our choice of relatively large (i.e., conservative) comoving 
distance cuts. As expected, the removal of nearby (i.e., the most strongly 
correlated) structure produces a prediction from the 2 point correlation 
function that is dominated by random/uncorrelated structure.

The angular separation distributions of the full redshift catalog and the 
zCOSMOS groups should be similar and increase monotonically with angular 
separation, which simply reflects 
the larger area on the sky sampled at larger angular separations. If, however, 
strong lensing clusters lie preferentially along lines of sight that are biased 
toward having more intervening structure as traced by our FoF groups then 
we expect the distribution of groups around strong lensing clusters to differ 
from that of the two control samples.

From Figure~\ref{fig:rdist} it is clear that we are seeing 
exactly this effect, in which the distributions of groups around our strong lensing 
clusters are significantly weighted toward small angular separations ($\theta$ 
$<$ 6\arcmin). This effect is apparent in the distributions using both the 30 and 
100 Mpc cuts to isolate uncorrelated structure. We can quantify the significance 
of the effect using the KS statistic -- specifically the two sided statistic as defined 
in \citet{press1992}. The distribution of groups around our strong lensing 
clusters is found to be inconsistent with the distribution of our input redshift 
catalog at 2.6$\sigma$ (2.4$\sigma$) for the 30 (100) Mpc cut to define 
uncorrelated structure. This strongly supports the hypothesis that the distribution 
of groups around our strong lensing clusters does not simply follow the distribution 
of redshifts in our spectroscopic catalog. We also find that the distribution of 
groups around our strong lensing clusters is inconsistent with the distribution of 
groups around the most massive structures in the zCOSMOS 10k catalog 
at 2.3$\sigma$ (2.2$\sigma$) for the 30 (100) Mpc cuts; this similarly supports 
the argument that our strong lensing selected lines of sight are strongly 
(i.e., $>$ 2$\sigma$) discrepant with clusters lying along random lines of sight. 
We also measure the KS statistic for our observed SL group distribution and 
the predicted distribution from the 2 point correlation function; the KS test results 
indicate that the strong lensing selected cluster represent the high end of the 
density of uncorrelated, project structure along the line of sight at the 2.6$\sigma$ 
(2.3$\sigma$) for the 30 (100) Mpc cuts.

\begin{deluxetable}{lccc}
\tablecaption{Groups Identified In Each Line of Sight\label{tab:groups}}
\tablewidth{0pt}
\tabletypesize{\tiny}
\tablehead{
\colhead{Cluster Name} &
\colhead{ \# Groups } &
\colhead{ $<$ \comdlimit~(100) } &
\colhead{ $>$ \comdlimit~(100) } \\
\colhead{ } &
\colhead{ } &
\colhead{ Mpc } &
\colhead{ Mpc }
}
\startdata
Abell 1703  &  7  &  1 (3)  &  6 (4)   \\
SDSS J0851+3331  &  3  &  0 (2)  & 3 (1)   \\
SDSS J0915+3826  & 11 &   5 (7)   & 6 (4)  \\
SDSS J0957+0509  &   7  &   0 (0)  & 7 (7)  \\
SDSS J1038+4849  &  10  &  4 (5)  &  6 (5)  \\
SDSS J1050+0017  &  17  &  5 (10)  & 12 (7)  \\
SDSS J1152+3313  &   8   &  0  (2) &  8 (6)  \\
SDSS J1226+2149\tablenotemark{a}  &  8  &  2 (5)  &  6 (3)  \\
SDSS J1226+2149   &   6  &   2 (2)  &  4 (4)
\enddata
\tablenotetext{a}{~Two strong lensing cluster cores are separated by $\sim$500 kpc 
and therefore occupy a single line of sight and share a spectroscopic dataset and 
the resulting group catalog.}
\label{tab:groupsummary}
\end{deluxetable}

%
%
%
%
%
%
%
%

Finally, we emphasize that our results are also limited by our spectroscopic 
sampling, which is relatively sparse in each cluster field compared to larger, 
dedicated spectroscopic surveys 
\citep[e.g.,][]{6dfref, sdssdr10,wigglezref,deep2ref,rines2013}. Even so, we 
detect a clear signal indicating that galaxy clusters that are selected for strong 
lensing preferentially lie along lines of sight that contain an over-abundance 
of projected structure relative to random lines of sight on the sky. Furthermore, 
as discussed above in $\S$~\ref{sec:selectioneffects}, the primary selection 
effect in our redshift data is the preferential targeting of photometrically-selected 
candidate cluster members which would only make it less likely that we identify 
groups with small angular projections relative to the strong lensing clusters. 
This effect biases our measurement low, and our measurement is therefore 
a lower limit on the degree to which line-of-sight structure is biased when 
looking toward strong lensing selected clusters are biased.

\section{Discussion}
\label{sec:discussion}

\subsection{Frequency of Line-of-Sight Structure: Comparison Against Semi-Analytic 
Predictions and LRG Over-Densities}

Recent studies have identified potentially powerful strong lensing lines of sight 
by presupposing that the cumulative effect of multiple massive structures along 
a single line of sight will produce large strong lensing cross sections \citep{wong2012, ammons2013,wong2013}. This is a novel approach to addressing the question 
of how important line-of-sight structure is to generating large strong lensing cross 
sections and provides an interesting point of comparison. In this paper we have 
approached the problem from the opposite direction, beginning with a sample 
of strong lensing lines of sight that are selected for high magnifications (i.e., 
sufficient to produce a giant arc that is visible in the SDSS), and large strong 
lensing cross-sections as indicated by the large Einstein radii of these lenses, 
all of which have $\theta_{E} >$ 5\arcsec (with eight of the nine having 
$\theta_{E} >$ 9\arcsec). 

We can compare our results against the expectation based on 
semi-analytic modeling; \citet{wong2012} argue that projected structures with 
angular separations of $\sim$100\arcsec~are optimal for maximizing lensing 
cross sections, and in \citet{wong2013} they measure structures as traced by 
luminous red galaxies (LRGs) within radial apertures of 210\arcsec. Our sample 
of strong lensing selected clusters have an excess of projected structure within this 
210\arcsec~aperture (Figure~\ref{fig:rdist}), and with the excess extending out 
to an aperture of $\sim$6\arcmin. Dedicated follow-up of two LRG-over-dense 
SDSS fields by \citet{ammons2013} shows that each field contains a single 
dominant cluster-scale structure as well as one or two additional groups lined 
up in projection with angular separations of $\sim$2-4\arcmin. These systems 
look very much like the cluster lenses in our sample that have groups in 
projection at small angular distances.


\subsection{Frequency of Line-of-Sight Structure: Comparison Against 
Predictions from Simulations}

Several studies have attempted to quantify the importance of the contribution 
of line-of-sight structure to the strong lensing cross sections of massive halos 
in cosmological simulations, drawing somewhat conflicting conclusions 
\citep{wambsganss2005,hilbert2007,puchwein2009}. By ray tracing to 
recover the strong lensing cross sections of massive halos in cosmological 
N-body simulations, \citet{wambsganss2005} find that $\sim$ 30-38\% of 
lensed sources occur because of contributions to the surface mass density 
from additional structures along the line of sight that are not physically 
associated with the primary lensing mass distribution.

\citet{hilbert2007}, on the other hand, perform a ray-tracing analysis of 
cluster-scale haloes in the Millennium Simulation and find that the mass 
attributed to line-of-sight structure toward cluster lenses contributes to the 
total surface mass density at only the few percent level. They conclude 
that mass associated with structures projected along the line of sight are 
modest and generally much smaller than found by \citet{wambsganss2005}. 
However, in a later analysis of strong lensing by clusters in the Millennium 
Simulation, \citet{puchwein2009} find that line-of-sight structure increases 
the total strong lensing optical depth by $\sim$ 10-25\%, and can frequently 
boost the strong lensing cross sections of individual clusters by as much as 
50\%. This result is much more in line with the findings of \citet{wambsganss2005}, 
and leaves us with a rather confusing picture of what simulations have to say 
regarding the role of line-of-sight structure in generating strong lensing 
galaxy clusters.

Specific predictions from simulations regarding increases 
in global strong lensing optical depth by galaxy clusters are not robustly 
testable using observations because we have no means of confidently 
attributing individual instances of strong lensing to the presence of line 
of sight structures. However, our results are qualitatively consistent with 
the claims of \citet{wambsganss2005} and \citet{puchwein2009} that line 
of sight structure has a significant impact on the generation of large strong 
lensing cross sections, and therefore on many instances of cluster-scale 
strong lensing.

\subsection{Implications for Strong Lensing Deep Fields and Cluster Cosmology}

Our observational results also have important implications regarding the 
efforts to reconstruct precision strong lens models and constrain the magnification 
of lensed background sources. \citet{daloisio2013} find that uncorrelated 
structure along the line of sight can often contribute fluctuations in the 
magnification of background sources at the $\sim$30\% level for magnifications 
of $\sim$ 10$\times$. Perturbations of this size are significantly larger than the 
typical uncertainty in the magnifications of strongly lensed sources that are 
estimated from high-fidelity strong lens models for cluster lenses 
\citep[e.g.,][]{kneib2011,sharon2012}. We have shown here that strong lensing 
galaxy clusters are biased toward lying along lines of sight with a large 
amount of intervening structure. It is therefore crucial that current and future 
efforts to use galaxy cluster lenses as cosmic telescopes account for the 
additional uncertainty in the strong lensing models due to line-of-sight structure. 
Non-parametric and hybrid methods for strong lensing reconstruction 
\citep[e.g.,][]{bradac2005a,coe2008,jullo2009} have the flexibility to allow 
for light deflection due to optically dark and/or uncorrelated line of sight 
structure. For cases where precision magnification maps are essential, it may 
be necessary to devote resources to characterizing structures along the line of 
sight and using observations -- such as velocity dispersion measurements of 
groups -- to inform parameter-based models of those  structures.

Our results have another important implication for efforts to constrain 
cosmological parameters by measuring the growth of structure as 
traced by the abundance of massive galaxy clusters as a function of mass 
and redshift (i.e., the cluster mass function). Current studies are limited by 
systematic uncertainties in measuring the masses of galaxy clusters 
\citep{benson2013,reichardt2013}. Weak lensing observations, in particular, 
are an important method for calibrating the normalization of other 
mass-observable relations  \citep[e.g., ][]{high2012}, but weak lensing 
measurements are plagued by large scatter in the mass estimates of 
individual clusters due in part to line-of-sight effects. For clusters along 
random lines of sight, projected large scale structure induces an 
additional uncertainty that is effectively random \citep[e.g.,][]{hoekstra2001}. 
However, any reasonably large cosmological survey for galaxy clusters will 
contain a subset of strong lensing galaxy clusters, and those strong lensing 
clusters are not likely to lie along random lines of sight. Rather, the strong 
lensing clusters will preferentially be those clusters that lie along lines of 
sight with an over-abundance of intervening structure (and conversely, the 
non-strong lensing clusters will preferentially lie along under-abundant 
lines of sight). This means that the impact on weak lensing measurements 
due to line-of-sight structure for these strong lensing clusters will not be to 
inject an additional random scatter, but rather to systematically bias the 
weak lensing measurements high. The presence of strong lensing in a 
subset of a cosmological cluster catalog can therefore be used as information 
to inform mass observable scaling relations, and probabilistically reduce the 
uncertainty in mass measurements of individual clusters.
%
%

\section{Summary and Conclusions}
\label{sec:conc}

We have presented the first measurement of the frequency of projected 
structures along the line of sight toward a sample of strong lensing selected 
galaxy clusters. There is clear an excess (relative to random lines of sight) of 
projected line-of-sight structure within small ($\leq$ 6\arcmin) angular 
apertures of our strong lensing selected cluster sample, and the distribution 
of structure around strong lensing groups is measured to be inconsistent 
relative to the expectation for a random line of sight at $\gtrsim$ 2.6$\sigma$. 
The small sample size of our strong lensing cluster sample and 
group catalogs are responsible for limiting the confidence of our measurement, 
but in spite of these limitations we still find strong evidence for line-of-sight 
bias toward strong lensing clusters in comparisons against two independent 
control samples.

Furthermore, these results have implications for astrophysical and 
cosmological observations involving galaxy clusters. Current and future 
studies that aim to make use of galaxy clusters lenses to magnify the 
background universe should consider and account for structure that is 
projected along the line of sight toward strong lensing galaxy clusters. 
There are also exciting possibilities for taking advantage of strong lensing 
information in future cosmological analyses of galaxy cluster samples. From 
the presence or lack of strong lensing by individual clusters it is possible 
to infer what clusters are likely to have excess cosmic structure projected 
along the line of sight, and that information may be used to reduce scatter 
in cluster mass measurements. 


\acknowledgments{We thank Eric Bell for helpful discussion that improved 
the quality of this paper significantly. This work was supported by the 
National Science Foundation through Grant AST-1009012. Support was 
also provided by NASA through grant HST-GO-13003.01 from the Space 
Telescope Science Institute, which is operated by the Association of 
Universities for Research in Astronomy, Incorporated, under NASA 
contract NAS5-26555, and by the FIRSTprogram ÒSubaru Measurements 
of Images and Redshifts (SuMIRe)Ó, World Premier International Research 
Center Initiative (WPI Initiative), MEXT, Japan, and 
Grant-in-Aid for Scientific Research from the JSPS (23740161). The data 
used in this work is publicly hosted with support from the Harvard-Smithsonian 
Center for Astrophysics (CfA) in collaboration with Harvard Library and the 
Institute for Quantitative Social Science, with infrastructure provided by 
Harvard University Information Technology Services}

\bibliographystyle{apj}
\bibliography{/Users/mbayliss/astro/master_bibliography}

\end{document}